\documentclass[twocolumn]{aastex62}
\usepackage[utf8]{inputenc}

\graphicspath{{fig4/}}
\usepackage{multirow}
\usepackage{graphicx}
\usepackage{txfonts}
\usepackage{natbib}

\revised{\today}
\usepackage{CJK}  

\begin{document}
\begin{CJK*}{UTF8}{gbsn}
\title{Unveiling the Hierarchical Structure of Open Star Clusters: the Perseus Double Cluster}

\correspondingauthor{Heng Yu, Zheng-Yi Shao}
\email{yuheng@bnu.edu.cn, zyshao@shao.ac.cn}

\author[0000-0001-8051-1465]{Heng Yu (余恒)}
\affil{Department of Astronomy, Beijing Normal University, Beijing, 100875, China.}

\author[0000-0001-8611-2465]{Zheng-Yi Shao (邵正义)}
\affil{Shanghai Astronomical Observatory, Chinese Academy of Sciences, 80 Nandan Road, Shanghai 200030, People's Republic of China.}
\affil{Key Lab for Astrophysics, 100 Guilin Road, Shanghai, 200234, People's Republic of China}

\author[0000-0002-4986-063X]{Antonaldo Diaferio}
\affiliation{Dipartimento di Fisica, Universit\`a di Torino, Via
  P. Giuria 1, I-10125 Torino, Italy}
\affiliation{Istituto Nazionale di Fisica Nucleare (INFN), Sezione di Torino, Via
  P. Giuria 1, I-10125 Torino, Italy}
  
\author[0000-0002-0880-3380]{Lu Li(李璐)}

\affiliation{Shanghai Astronomical Observatory, Chinese Academy of Sciences, 80 Nandan Road, Shanghai 200030, People's Republic of China.}
\affiliation{University of the Chinese Academy of Sciences, No.19A Yuquan Road, Beijing 100049, China}

\begin{abstract}

We introduce a new kinematic method to investigate the structure of open star clusters. We adopt a hierarchical clustering algorithm that uses the celestial coordinates and the proper motions of the stars in the field of view of the cluster to estimate 
	a proxy of the pairwise binding energy of the stars and arrange them in a binary tree. 
The cluster substructures and their members are identified
by trimming the tree at two thresholds, according to the $\sigma$-plateau method. Testing the algorithm on 100 mock catalogs shows
that, on average, the membership of the identified clusters is $(91.5\pm 3.5)$\% complete and the fraction of unrelated stars is $(10.4\pm 2.0)$\%.
We apply the algorithm to the stars in the field of view of the Perseus double cluster from the Data Release 2 of Gaia. This approach
	identifies a single structure, Sub1, that separates into two substructures, Sub1-1 and Sub1-2. These substructures coincide with $h$ Per and $\chi$ Per: 
 the distributions of the proper motions and the color-magnitude diagrams of the members of Sub1-1 and Sub1-2  are fully consistent
	with those of $h$ Per and $\chi$ Per reported in the literature. These results suggest that our hierarchical clustering algorithm can be a powerful tool to unveil the complex kinematic information of star clusters.

\end{abstract}


\section{Introduction}
An open cluster is a group of stars which formed within the same giant molecular cloud and were roughly born at the same time. An open cluster generally is loosely gravitationally bound: some of the stars can leave the group  after their birth, while others can be tidally removed by close encounters with gas clouds and other star systems.
Open clusters are unique laboratories of stellar evolution, and also important probes of the structure 
and evolution of the Galactic disk. 

The identification of the  cluster members is the first crucial step for their investigation.
Open clusters do not usually show a large density contrast on the sky, unlike globular clusters; therefore,
sophisticated algorithms are necessary to identify their star members.
The method of the maximum likelihood pioneered by \citet{1958Vasilevskis} and \citet{1971Sanders}, based on bivariate Gaussian distributions of the proper motions of the cluster and field stars, 
has been revised and updated by several more recent studies \citep[e.g.,][to mention a few]{1990Zhao,1995Kozhurina-Platais,2004Deacon,2004Kharchenko,2006Dias,2010Krone-Martins,2014Sarro,2016Sampedro}, 
and remains the most adopted method. 

To avoid the bias introduced by a fixed density distribution, some non-parametric approaches have also been 
developed \citep[e.g.,][]{1990Cabrera,2004Balaguer,2006Javakhishvili,2019Nambiar}.
In addition, clustering methods have recently been introduced in the field.
\citet{2011Schmeja} explores four clustering  algorithms applied to the two-dimensional distribution 
of the stars on the sky, thus ignoring any kinematic information: star counts, nearest-neighbor density,  Voronoi tessellation, and the minimum spanning tree. \citet{2011Schmeja} concludes that the nearest-neighbor density is the most reliable method.

The Gaia mission provides unprecedented astrometric, photometric 
and spectroscopic data of stars and star clusters \citep{2016Gaia, 2018Gaia, 2018Gaiadr2}, that are ideal to test standard and new clustering algorithms.

\citet{2014Krone} combined the principal component analysis and the $k$-means clustering to design the UPMASK method, that
has been applied by \citet{GAIA2} to the data of the {\it Gaia} data release 2 (DR2),   thus including the stellar kinematic information. \citet{GAIA2} identify the members of 1229 star clusters, including $h$ Per and $\chi$ Per. 
An additional clustering algorithm, DBSCAN, based on the local density of points in some parameter space \citep{DBSCAN}, was applied by \citet{2014Gao} to the stars in the field of NGC188 with known proper motions and radial velocities.  \citet{2018Castro} also investigate the optimal parameters of 
DBSCAN on simulated data. 

Some of the methods mentioned above do not actually distinguish between members and non-members of the cluster, but rather assign a membership probability to each star in the field.  
Here, we propose a hierarchical clustering method, a new kinematic method based on  a simple physical quantity: a proxy for the pairwise gravitational binding
energy. This method arranges the stars in the field of view in a binary tree; by trimming this tree
according to the $\sigma$-plateau method \citep{1999Diaferio, Serra2011}, the method separates the star distribution into structures with unambiguously identified star members. 

Hierarchical clustering algorithms are well-known in computer science and statistics. They separate a system into subgroups based on the measure of an adopted similarity or metric \citep[see, e.g.,][for a detailed description]{everitt2011}. A hierarchical clustering algorithm was adopted by \citet{1978Materne} and \citet{Serna1996} to investigate groups and clusters of galaxies. 
\citet{1999Diaferio} and \citet{Serra2011} improved over the original algorithm by introducing the $\sigma-$plateau criterion to 
identify both the galaxies that are members of a galaxy cluster \citep{2013Serra} and the cluster 
substructures \citep{Yu2015, 2016Yu, Liu2018b}.  
In principle, this method can also be appropriate for other systems held together by gravity, like star clusters. Here, we show that this is indeed
the case and apply the method to the Perseus double cluster.

The Perseus double cluster is a bright and rich  open cluster, located at the distance of 2344$^{+88}_{-85}$ pc from the Sun \citep{2002Dias,2010Currie,gaia2018b}, with a relatively young age of about $\sim 12.6-14$~Myr \citep{2001Keller,2010Currie}.
Based on different data sets and methods \citep{2002Uribe,2010Currie,2013Kharchenko,gaia2018b,GAIA2}, 
many of the Perseus properties have been extensively investigated, including the stellar mass function \citep{2002Slesnick}, the mass segregation \citep{2005Bragg}, the substructures and its surrounding stellar halo \citep{2010Currie, 2019Zhong}, 
and the extended main-sequence turnoff \citep{2019Li}. The two main components of the Perseus cluster, $h$ Per (NGC869) and $\chi$ Per (NGC884), have similar photometric and spectroscopic properties.
It follows that separating their members with methods based on photometric data alone is not a trivial task.
The two Perseus components are clearly separated on the sky, so  an usual and simple strategy is to consider a star
as a member of one of the two components if it is located within one of two areas of the sky chosen {\it a priori}.

The hierarchical clustering method we propose here  identifies the substructures in the field of view of the cluster without assuming the position and size of the substructures in advance.
The precise measurement of the {\it Gaia} DR2 data of the Perseus cluster provides an ideal test of our method.

We describe the method in Sect. \ref{sec:method} and test it on mock catalogs in Sect. \ref{sec:simu}. We apply the method on the data in the field of view of Perseus 
 in Sect. \ref{sec:data}, and we present our results in Sect. \ref{sec:result}. We conclude 
in Sect. \ref{sec:end}.

\section{Method}
\label{sec:method}
Our hierarchical clustering algorithm arranges all the objects in the field of view in a binary tree 
according to a proxy of the pairwise gravitational binding energies of the objects: pairs of objects with increasingly absolute value of the binding energy 
will appear at increasingly deeper levels of the tree. By trimming the binary tree at appropriate thresholds, we can associate the tree branches to well-defined kinematic structures. We apply this procedure to the star cluster.

\subsection{The binary tree} 

The pairwise binding energy $E_{ij}$ of any pair of stars $i$ and $j$ combines their 
gravitational potential energy and their kinetic energy: 
\begin{equation}
E_{ij}=-{Gm_im_j\over r_{ij}} + {m_im_j\over 2(m_i+m_j)}v_{ij}^2\, ,
\end{equation} 
where $m_i$ and $m_j$ are the star masses and $r_{ij}$ and $v_{ij}$ are the pairwise relative distance and velocity, respectively, with $G$ the gravitational
constant.  In principle, we could estimate $E_{ij}$ by knowing the mass of the two stars and their six phase-space coordinates. 

Here, we intend to apply the algorithm to the field of view of Perseus. Therefore, for the three spatial coordinates, we consider the two celestial coordinates alone and ignore the distance of the star from the observer. In fact, the 
average uncertainty on the star parallax is $\sim 0.1$ mas, that corresponds to an uncertainty  of $\sim 500$~pc for the distance to
a star within the Perseus cluster;
it follows that the uncertainty on the star distance is $\sim 10$ times larger than the cluster size $\sim 41$~pc. We thus assume that all the stars are at the same
distance, corresponding to the distance $r=2.34$~kpc of the star cluster. 

As for the velocity components, in {\it Gaia} DR2 
the typical uncertainty on the proper motion  is $\sim  0.2$ mas yr$^{-1}$ for a
star of brightness $G = 17$~mag \citep{2018Gaiadr2}, which corresponds to a velocity of $\sim 2$ km s$^{-1}$ at the Perseus distance  $\sim 2$~kpc.  As we will see below, this uncertainty is comparable to the widths of the distributions
of the proper motions: the uncertainty is thus comparable to the velocity dispersion of the star cluster. 
Nevertheless,  we include the proper motions in the estimation of the pairwise binding energy. On the contrary, the uncertainties of the radial velocities are much larger than the uncertainties of  the proper motions for stars of brightness $G = 17$~mag.  
In fact, an uncertainty as small as $\sim 1$ km s$^{-1}$ on the radial velocity can only be obtained for much 
brighter stars, with $G = 12$~mag \citep{2018Gaiadr2}. We therefore ignore the stellar radial velocities. 

In conclusion, we consider only four out of the six phase-space coordinates, due to the large uncertainties
of the two neglected coordinates.  In Sect. \ref{sec:simu}, we perform a test that shows that these four coordinates are indeed sufficient to provide an appropriate proxy of the binding energy.

We thus estimate the pairwise binding energy of each star pair as 
\begin{equation}
{E_{ij}=-G{m_i m_j\over r~\theta_{ij}}p+{1\over 2}{m_i m_j\over m_i+m_j}\frac{(\Delta \mu_x^2 + \Delta \mu_y^2)}{2}r^2 }\;,
\label{eq:pairwise-energy}
\end{equation}
where $r=2344$~pc is the adopted distance to the Perseus cluster, 
$\theta_{ij}$ is the pairwise projected angular separation  of each star pair,
$\Delta \mu_x$ and $\Delta\mu_y$ are the pairwise differences of the two orthogonal components of the proper motion, 
$\mu_x = \mu_{RA} \cos\delta$ and 
$\mu_y=\mu_{DEC}$, with $\mu_{RA}$ and $\mu_{DEC}$ the proper motions along the celestial coordinates
right ascension $\alpha$ and declination $\delta$.\footnote{Neglecting the component along the line of sight of both the position and the velocity of each star clearly overweights the gravitational potential energy over the kinetic energy, because we underestimate both the three-dimensional separation and the three-dimensional relative velocity. This effect is easy to quantify in spherically symmetric systems: for a  three-dimensional pairwise separation $r_{ij}$ 
and an angle $\chi$ between $r_{ij}$ and the line of sight, we have the projected  pairwise separation  $\hat r_{ij}=r_{ij}\sin\chi$;  
by averaging over all the possible lines of sight, we find $\langle 1/\hat r_{ij}\rangle = 1/r_{ij}  \langle 1/\sin\chi\rangle =  \pi/(2 r_{ij})  $. 
Similarly, for the pairwise kinetic energy per unit mass $v_{ij}^2$, we find the average projected pairwise kinetic energy per unit mass $\langle \hat v_{ij}^2\rangle = 
 v_{ij}^2 \langle \sin^2\chi\rangle = (2/3) v_{ij}^2$. Therefore, we overestimate the absolute value of the gravitational potential energy by a factor $\pi/2$ and
 underestimate the kinetic energy by a factor $3/2$.   }

 Assigning the masses $m_i$ and $m_j$ is  a crucial issue that determines the correct balance between the gravitational and kinetic energy contributions. The simplest approach to assign the mass to each star could rely on the mass-luminosity relation. Unfortunately, this method is prone to be dominated by the uncertainty on the  distance of the stars and their derived luminosity: foreground bright stars erroneously associated to the cluster can generate spurious gravitational potential wells and background faint stars can erroneously be associated to cluster substructures.  
We thus set $m_i=m_j=m$ for any $i$ and $j$ to
avoid unnecessary complications deriving from 
these uncertainties, and set $m=1$~M$_\odot$;  this mass appears to be a reasonable value according to the recent mass function of open clusters \citep{2010Bastian}.  

To control the balance between the two energy contributions in eq. (\ref{eq:pairwise-energy}), we introduce the parameter $p$, that is automatically 
determined by our algorithm, as we illustrate in Sect. \ref{sec:pvalue} below. 
As mentioned above, $v_{ij}$ is affected by a large uncertainty which is comparable to the velocity dispersion of
the star cluster, and overestimating  $v_{ij}^2$, and thus the kinetic energy contribution, is thus more likely than underestimating it, compared to a situation where more accurate
measures of $v_{ij}$ were available. To restore the correct balance between the gravitational and kinetic energy contributions, it appear thus reasonable to underweight
the kinetic energy contribution by artificially amplifying the gravitational energy; we obtain this result by introducing the factor $p$, which will always be larger than 1. 
The pairwise binding energy
$E_{ij}$ will not correspond to its actual value, but we are
more likely to preserve the correct relative weight of the two
energy contributions.

For a catalog of $N$ stars, the binary tree is built as follows (see \citealt{1999Diaferio,Serra2011,Yu2015} for further details): 

\begin{itemize}
	\item[i.] initially, each star $\alpha$ is an individual group $G_\alpha$, and we thus have $N$ groups $G_i$, with $ {i=\{1,\dots,N\}}$; 
\item[ii.] we estimate the binding energy between two groups $G_\alpha$ and $G_\beta$ as $E_{\alpha\beta}={\rm min}\{E_{ij}\}$,
	where $E_{ij}$ is the pairwise binding energy between the star $i\in G_\alpha$ and the star $j\in G_\beta$; 
\item[iii.] the two groups with the smallest binding energy are
replaced with a single group and the total number of groups is decreased by one;
\item[iv.] we repeat the procedure from step (ii) until only one group is left.
\end{itemize}

\begin{figure*}[htp]
\includegraphics[width=0.95\textwidth]{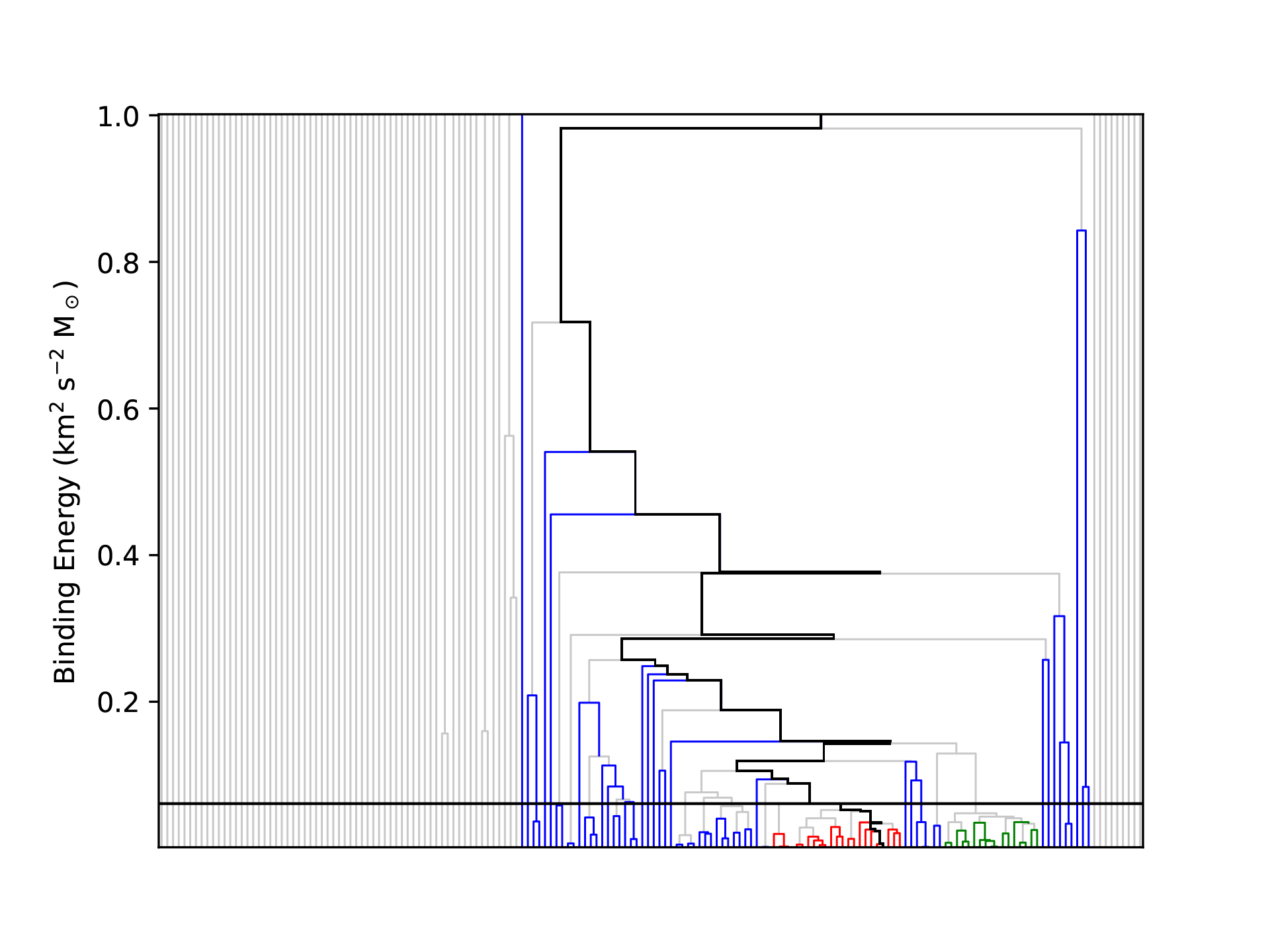} \\
\caption{Dendrogram of the 172 stars brighter than $G=10$ mag in the field of view of Perseus. 
The stars are the leaves of the tree at the bottom of the dendrogram. 
The vertical coordinate of each node is its binding energy. 
The black path highlights the main branch. The horizontal black line 
shows the lower threshold; the upper threshold is closer to the root and is outside the plot.
The upper threshold identifies the main structure whose leaves are highlighted by the blue 
branches from which they hang; the lower threshold identifies two substructures whose leaves are highlighted by the red and green  branches from which they hang.}
\label{fig:tree0}
\end{figure*}

At the end of the procedure, all the stars in the field of view are arranged in a binary tree. As an example, Fig. \ref{fig:tree0} 
shows the binary tree obtained by applying the algorithm to the 172 stars brighter than $G=10$ mag in the field of view of Perseus. For this illustrative example, we set $p=1$.

The graphical representation of the binary tree is called {\it dendrogram}.
Each segment shown in the figure is a tree branch that links two nodes. The two nodes hanging below each node, which is the parent node, are called children. The root
at the top of the dendrogram, which is not shown in Fig. \ref{fig:tree0}, is the parent of all the nodes. The nodes without children, at the bottom of the dendrogram, 
are the leaves. Each star is a leaf at the bottom of the dendrogram, and each group of stars is identified by each node at each level
of the dendrogram. 
The ordinate of each node is its binding energy, whereas its abscissa value is set to  
properly display the dendrogram.

\subsection{Trimming the tree}
\label{sec:tree}

The identification of the stellar structures in the tree requires the definition of a threshold 
to trim the branches of the tree. To set this threshold, we consider that a gravitationally bound cluster can be roughly approximated by an isothermal sphere;
therefore, different subsamples of the cluster members should approximately return the same estimate of the velocity dispersion of the cluster \citep{1999Diaferio,Serra2011}.

\begin{figure}[htp]
\includegraphics[width=0.45\textwidth]{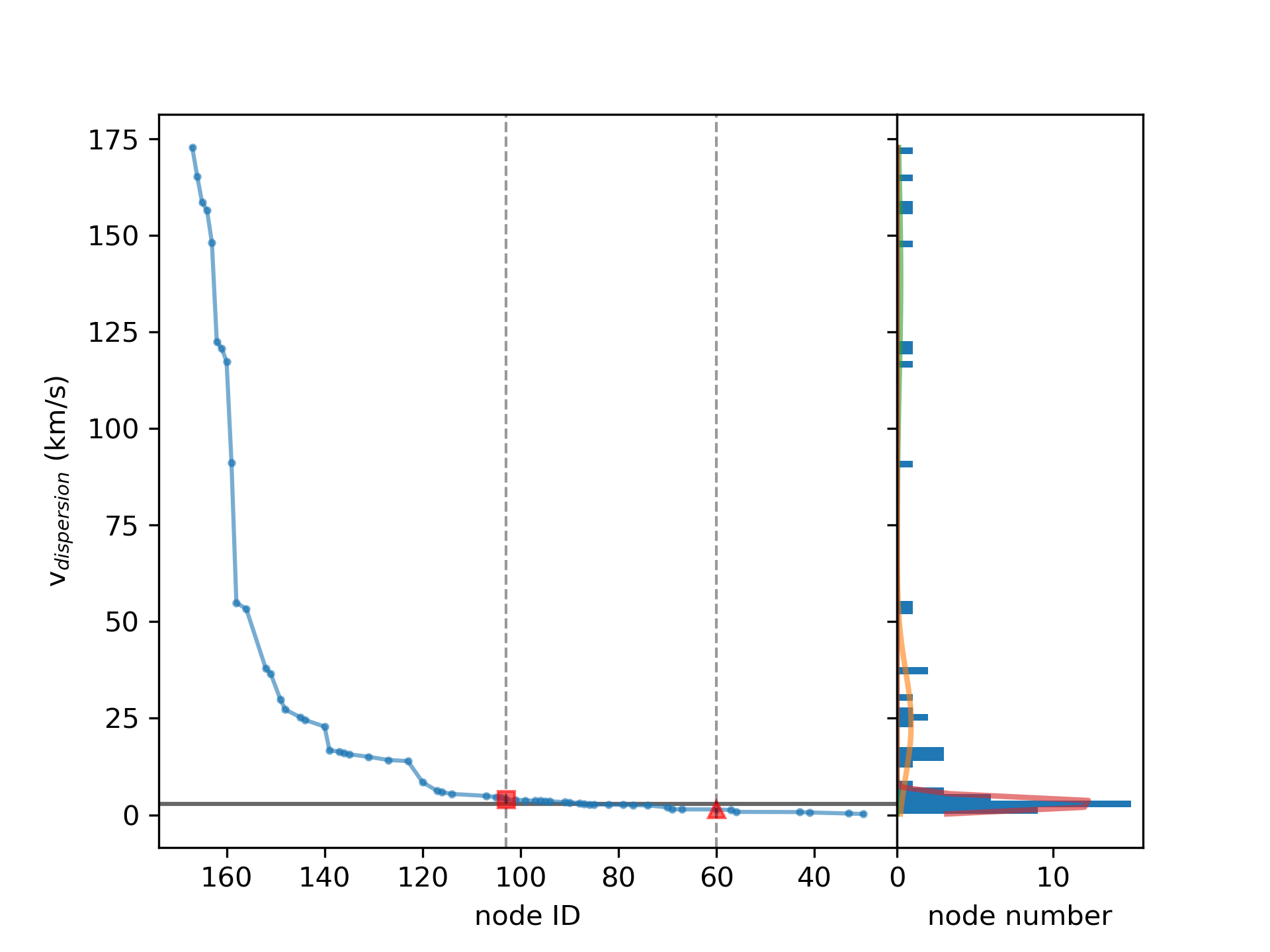} \\
\caption{Velocity dispersion of the leaves hanging from the nodes on the main branch of the dendrogram shown in Fig. \ref{fig:tree0} as a function of the node identification number. The root of the binary tree is on the left; the leaves are on the right. The $\sigma$ plateau is shown by the horizontal solid line; the vertical dashed lines show the plateau extension. The red square and triangle indicate the key nodes of the plateau. The distribution of the velocity dispersions on the main branch nodes is shown in the right panel. Its multi-Gaussian fit returns three components shown by the red, orange, and green curves.}
\label{fig:plateau}
\end{figure}

We can exploit this feature when the cluster we are interested in is the richest system in the field of view. In this case, the cluster corresponds to the main branch of the tree, namely the set of nodes, 
at each level of the tree, from which the largest number of leaves hangs. In Fig. \ref{fig:tree0}, the main
branch is highlighted by the thick black line. 
We walk along the main branch and compute the velocity dispersion $\sigma$ of the leaves hanging from each node on the main branch at each level of the tree. Figure \ref{fig:plateau} shows  $\sigma$ on the main branch of the binary tree of Fig. \ref{fig:tree0} as a function of the main branch
nodes. The node identification numbers on the horizontal axis are sorted from the root on the left to the leaves on the right of the panel.
This figure shows that when walking from the root to the leaves, the velocity dispersion drops rapidly, reaches a long plateau, and decreases again. This latter drop is not actually obvious for the small sample we show in Fig. \ref{fig:plateau}, but it appears more clear in richer structures, as for the structure shown in  Fig. \ref{fig:mb} below. We call this plateau the $\sigma$ plateau.

To locate the $\sigma$ plateau, we consider the distribution of $\sigma$, as shown in the right panel of Fig.  \ref{fig:plateau}. We fit this distribution with the Gaussian mixture model \citep[GMM,][]{scikit-learn}
\begin{equation}
 G = \sum_i^{N_G} \; w_i \; g(\sigma_i,\delta_i) = \sum_i^{N_G} \; \frac{w_i}{\delta_i\sqrt{2\pi}} e^{-(x-\sigma_i)^2/2\delta_i^2}\, ,
\end{equation}
where the sum is over the $N_G$ Gaussians $g(\sigma_i,\delta_i)$. $N_G$ is set according to the Bayesian Information Criterion \citep[BIC;][]{ivezic2014statistics}, that selects the 
model with the smallest BIC to minimize the number of free parameters required to fit the data. We limit the search of $N_G$ in the range $1-10$.
For each Gaussian component, the fit returns its mean $\sigma_i$, its standard deviation $\delta_i$ and its weight $w_i$. 
The largest $w_i$ identifies the principal Gaussian component, and thus its parameters $\sigma_0$ and $\delta_0$. The principal Gaussian component captures the $\sigma$ plateau. 
We identify the $\sigma$ plateau from $\sigma_0$ and $\delta_0$:  
the nodes corresponding to $\sigma_0 + \delta_0$ and $\sigma_0 - \delta_0$  define the extension of the plateau; these nodes are highlighted by the two vertical dashed lines in 
Fig. \ref{fig:plateau}. 
The plateau generally is neither exactly flat nor monotonically decreasing; we thus identify the nodes associated with the largest and the smallest velocity dispersions within the plateau extension and
call them the key nodes. The key nodes are highlighted with the red symbols in Fig. \ref{fig:plateau}. Here, the key nodes coincide with the plateau extension, but this might
not always be the case, as it happens for the case shown in Fig. \ref{fig:mb}. 
We adopt the binding energies of these two key nodes as the thresholds for trimming the tree: the leftmost and rightmost key nodes identify the cluster and its substructures, respectively. 

\subsection{Setting the parameter $p$}
\label{sec:pvalue}
 The value of $p$ is identified by requiring an appropriate balance between the gravitational and the kinetic
energy contributions in eq. (\ref{eq:pairwise-energy}): 
if $p$ is too small, the gravitational energy is underestimated,
a substantial number of real members are not associated to the cluster and the extension of the $\sigma$ plateau is underestimated;
if $p$ is too large, the gravitational energy is overestimated and
a substantial number of interlopers, erroneously associated to the cluster, blur the appearence of the $\sigma$ plateau.  

We identify the proper value of $p$ by exploiting its effect on the weight $w_0$ of the
principal Gaussian component mentioned in Sect. \ref{sec:tree}: the largest possible $w_0$ identifies the optimal $p$. We adopt the three-point equal-interval search scheme \citep{ravindran2006}.  
We start with three values of $p=\{p_0,p_1,p_2\}=\{1, 20, 40\}$. For each of them, we derive the binding energies and build the binary tree. We thus derive the corresponding value of $w_0$ for each $p_i$. 
If $w_0$ increases monotonically with $p_i$, we repeat the procedure for the fourth value of $p$, $p_3=2p_2 - p_1 =60$. 
We repeatedly do so until $w_0$ decreases. 

We thus find the maximum $w_0$ corresponding to 
the value $p_M$ of our set $\{p_0, p_1, \dots, p_{M-1}, p_M, p_{M+1}\}$, with $p_{i+1}=2p_i - p_{i-1}$, for $i\ge 2$. 
The values $p_{M-1}$ and $p_{M+1}$ define the length $L=p_{M+1}-p_{M-1}$.
We now derive $w_0$ for $p_{\mathrm {left}}=p_{M-1}+L/4$ and $p_{\mathrm{right}}=p_{M-1}+3L/4$
and identify the largest $w_0$ among the five values corresponding to the set of $p$, $\{p_{M-1}, p_{\mathrm {left}},p_M,p_{\mathrm{right}}, p_{M+1}\}$. We thus identify the new set of three consecutive values of $p$, $\{p_a,p_b,p_c\}$, where the largest $w_0$ is associated to $p_b$. This set, which has now length $L^\prime=L/2$, is taken as the new searching range and we again derive $w_0$ for  $p_{\mathrm {left}}^\prime=p_a+L^\prime/4$ and $p_{\mathrm{right}}^\prime=p_a+3L^\prime/4$. 
We iterate this procedure until the length of the searching range $L$ is smaller than 3.

\begin{figure}[htp]
 \centering
\includegraphics[width=0.45\textwidth]{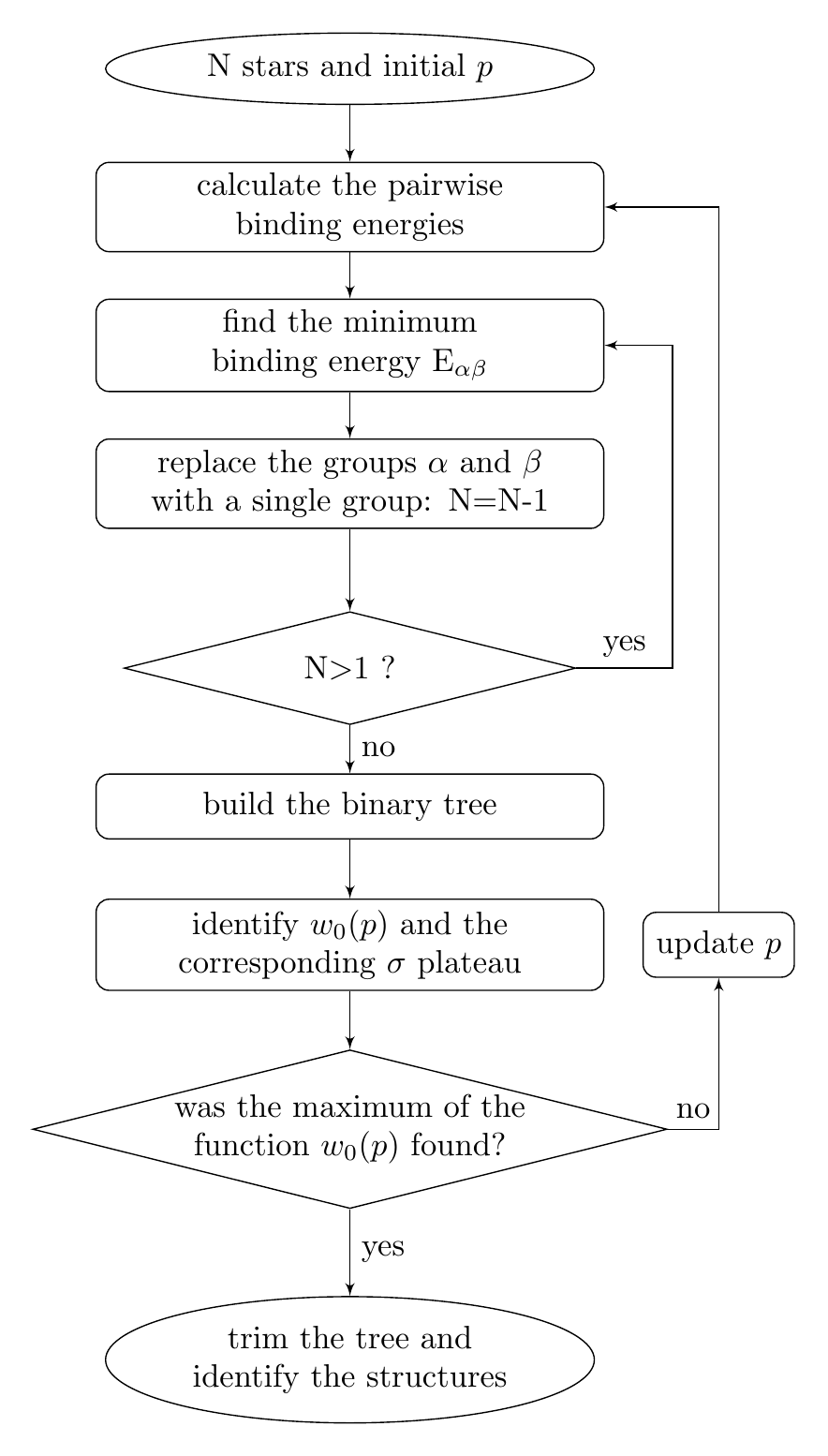}
\caption{ Flow chart of the procedure for the identification of the cluster structures.}
\label{fig:flowchart}
\end{figure}

 Figure \ref{fig:flowchart} shows a flow chart of our algorithm. The core of the procedure is the classical hierarchical clustering method, that we implemented based on the python module {\it scikit-learn} \citep{scikit-learn}. The crucial modifications we brought to this module are the adopted similarity and the trimming criterion: for the former, we replace the original dimensionless distance with the proxy $E_{ij}$ (eq. \ref{eq:pairwise-energy}) of the gravitational binding energy; for the latter, the criterion is based on the $\sigma$ plateau method described in the previous section, that identifies
 both the cluster and its substructures. 
The most time-consuming section of the algorithm is the calculation of the pairwise binding energy and 
thus the computational cost of the algorithm is proportional to $N^2$, with $N$ the number of stars in the field. 

\section{Mock catalogs}
\label{sec:simu}
 We test our method on a set of mock catalogs. 
We build a synthetic cluster of 3000 members whose number density distribution on the sky follows the spherical King's model \citep{1962King}:
\begin{equation}
\rho(r)=\rho_{\rm c} \left( {1\over \sqrt{1+r^2/r_{\rm c}^2}} - {1\over \sqrt{1+r_{\rm t}^2/r_{\rm c}^2}} \right)^2\, .
\end{equation}
We set the tidal radius r$_{\rm t}$ to infinity and derive the other parameters of the model to roughly mimic the Perseus cluster at its distance $\sim 2.5$~kpc.  
The core density $\rho_{\rm c}=8.9$~arcmin$^{-2}$ returns $\sim 3000$ members within a circular region of radius 48~arcmin; this number of members is comparable to $\sim 2500$, the total number of members  of NGC884 and NGC869 combined, according to \citet{GAIA2}. To mimic the Perseus cluster as a whole, we adopt the core size $r_{\rm c}=5$~arcmin, a value larger than the core radius of NGC884 or NGC869 separately, $r_{\rm c}\sim 2$~arcmin. 
We simulate the proper motions and velocity dispersion of the Perseus stars by sampling the proper motion components of the mock stars from  Gaussian distributions with means and standard deviations $\mu_x  = -0.65 \pm 0.18$~mas~yr$^{-1}$ and $\mu_y =-1.05 \pm 0.18$~mas~yr$^{-1}$, as estimated from the GAIA DR2 data \citep{2019Li}. The parallaxes of the stars
are sampled from a Gaussian distribution with  mean and standard deviation $0.40 \pm 0.06$~mas. 
The standard deviation is the instrumental error of GAIA \citep{2018Gaiadr2}, because the error associated to the size of the cluster is negligible. The line-of-sight velocities are sampled from a Gaussian distribution with mean and standard deviation $-45.0 \pm 2.1$~km~s$^{-1}$, according to the recent measurement of NGC884 \citep{gaia2018b}.

As background stars, we select the 29446 stars brighter than $G=18$~mag from the circular region of radius 48~arcmin located one degree south of NGC869, where no cluster exists. We keep all the properties of these background stars unaltered, namely celestial coordinates, proper motions and parallaxes.
For $\sim 97$\% of these stars, the radial velocity measures are unavailable. To the stars with missing radial velocity we associate a radial velocity by sampling the Gaussian distribution of the available measures, whose mean and standard deviation are $-32\pm 33$~km~s$^{-1}$. 

We now assign the celestial coordinates to the stars of the mock cluster so that we have the cluster at the center of this field of background stars.
We randomly select 7000 stars from the background stars and create a catalog of 10000 stars, so that  30\% of the catalog stars are cluster members. 

 From this mock catalog, we extract 20 subsamples with $N$ randomly selected stars. We adopt 5 values of $N= 500, 1000, 2000, 3000, 4000$. We thus end up with 100 different mock catalogs. 
To limit the computing time of our algorithm, we stop the iteration for the identification of $p$ when the range $L$,  described in the previous section, drops below 10 rather than 3. 

\begin{figure}[htp]
\centering
\includegraphics[width=0.45\textwidth]{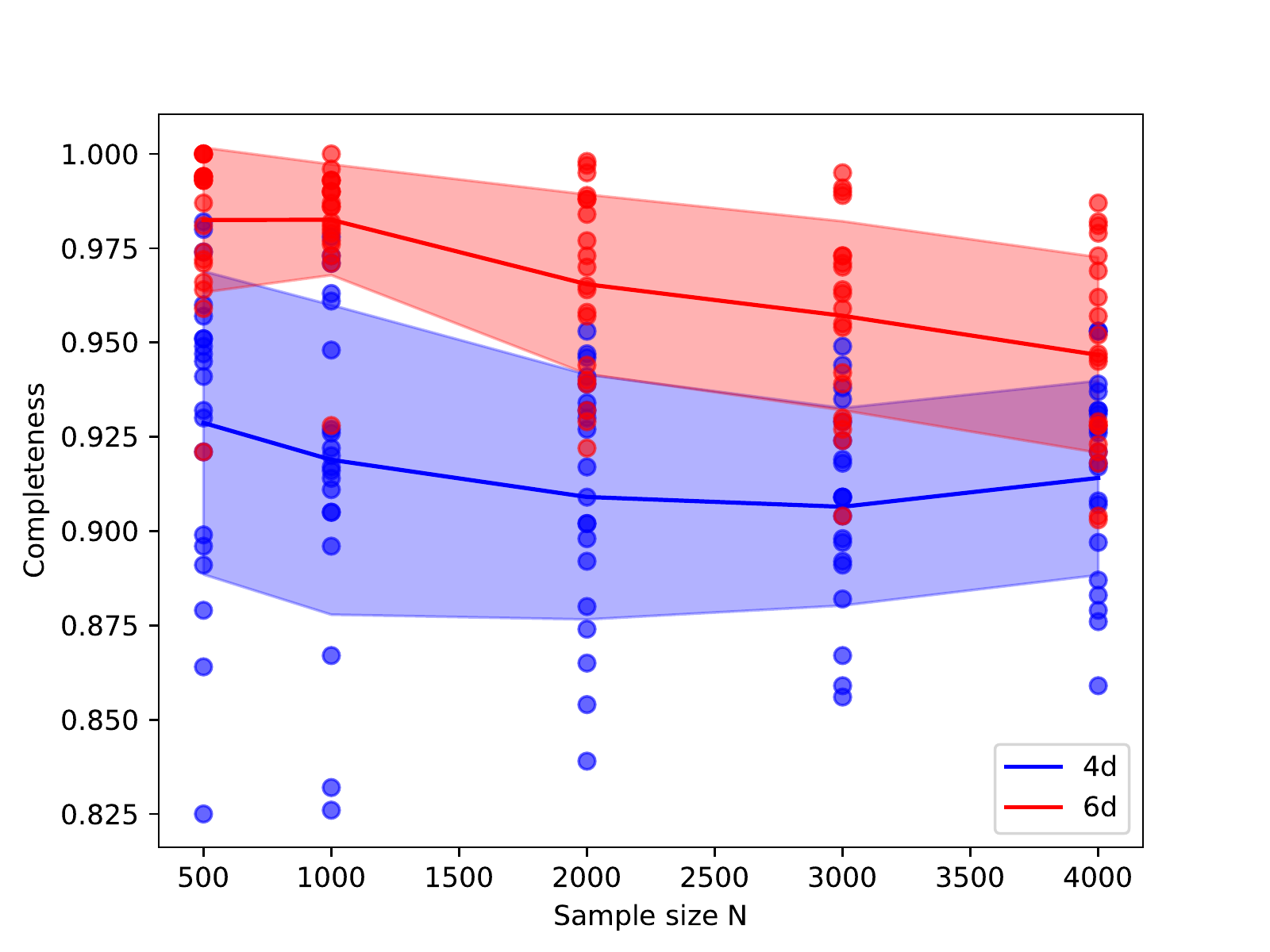}
\caption{ Completeness of the main structure in the 100 mock catalogs as a function of the size $N$ of the sample. 
The blue (red) dots show the completeness from the analysis with four (six) phase-space coordinates alone (eqs. \ref{eq:pairwise-energy} and \ref{eq:pairwise-energy6}, respectively). The curves and shaded areas show the mean and standard deviation of each sample at fixed $N$. }
\label{fig:sdist}
\end{figure}

\begin{figure}[htp]
\centering
\includegraphics[width=0.45\textwidth]{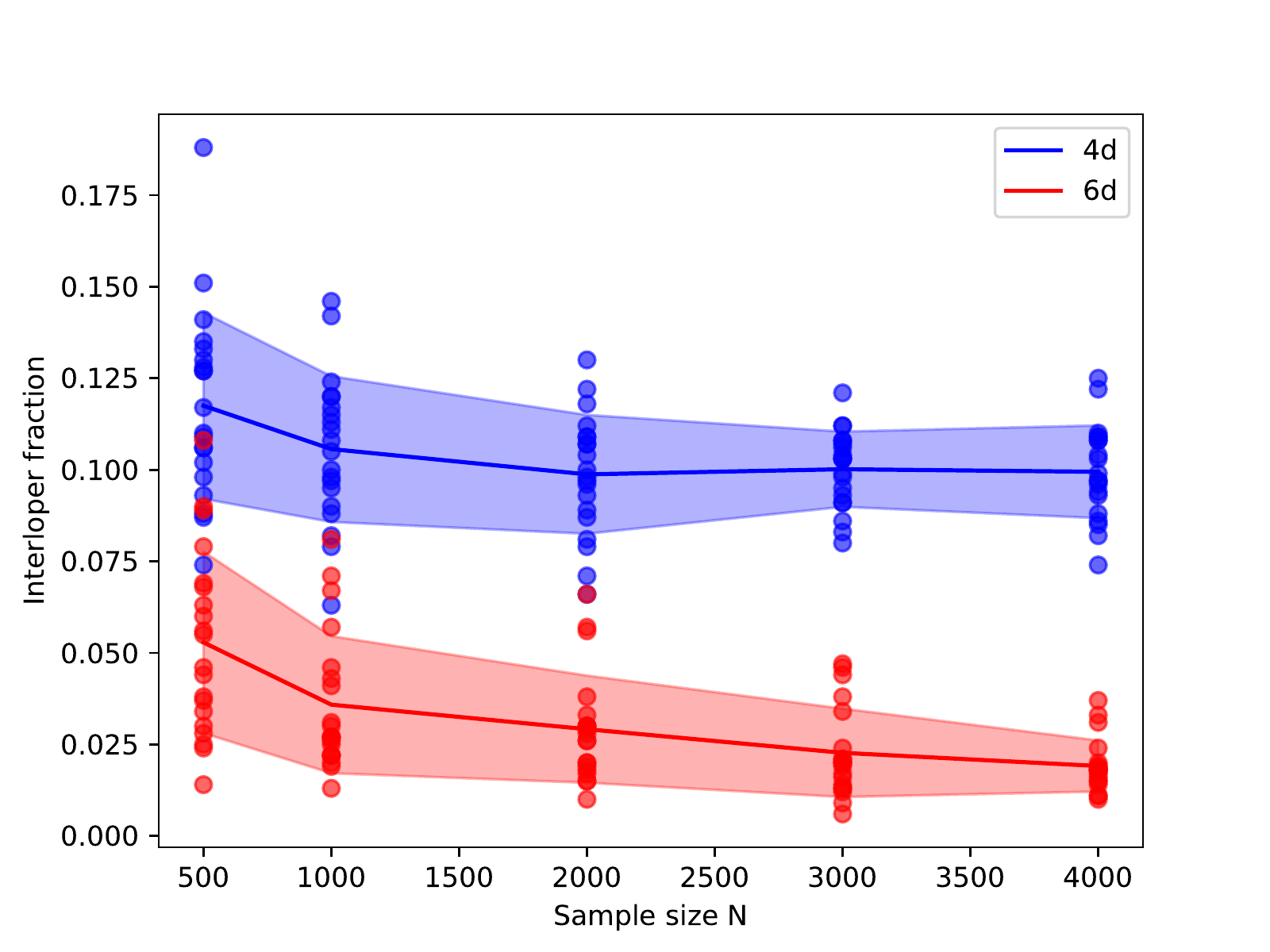}
\caption{ Same as Fig. \ref{fig:sdist} for the fraction of interlopers.}
\label{fig:fdist}
\end{figure}

The blue dots in Fig. \ref{fig:sdist} show the completeness of the catalogs, namely the fraction of the members of the identified main structure that actually are real members, as a function of the sample size $N$. The blue dots of Fig. \ref{fig:fdist} shows the interloper fractions, namely the fraction of the members of the identified  main structure that actually are background or foreground stars, as a function of $N$.  The completeness and the interloper fractions appear basically independent of the sample size $N$, for the range we investigate here. The average completeness and interloper fractions are
$(91.5 \pm 3.5)$\% and $(10.4 \pm 2.0)$\%, respectively.

 In the estimation of the binding energy we only include the two celestial coordinates and the two proper motion components. To quantify the systematic error caused by ignoring the distance to the star and its radial velocity,  we also compute the completeness and the interloper fraction when we adopt the full expression for the binding energy
\begin{eqnarray}
E_{ij} &=& -~G{m_i m_j\over r_i^2 + r_j^2 - 2r_i r_j {\rm cos}\theta_{ij}}p \\
  &+& {1\over 2}{m_i m_j\over m_i+m_j} \frac{ (\mu_{xi}r_i-\mu_{xj}r_j)^2  + (\mu_{yi}r_i-\mu_{yj}r_j)^2 + (v_i - v_j)^2 }{3} ,  \nonumber 
\label{eq:pairwise-energy6}  
\end{eqnarray}
with obvious meaning of the symbols. In this expression we still need to include the parameter $p$, because we continue to assume $m_i=m_j=1$~$M_\odot$.
As in the previous tests, to limit the computational effort, we stop the procedure for the identification of $p$  when the searching range $L$ drops below 10.

With the six phase-space coordinates, the mean completeness increases to $(96.7 \pm 2.6)$\%, as shown by the red dots in Fig. \ref{fig:sdist}, whereas the fraction of interlopers drops to $(3.2 \pm 2.0)$\%, as shown by the red dots in Fig. \ref{fig:fdist}.
Despite the improved results, the corresponding completenesses obtained with the four phase-space coordinates alone are well within $2\sigma$ from the six-coordinate values. The bias on the interloper fractions is relatively larger, but still within $3\sigma$ from the six-coordinate values. We thus 
conclude that limiting our algorithm to four phase-space coordinates does return a biased completeness and a biased interloper fraction compared to an approach where all the six coordinates were known; however, these biases are expected to be within the random fluctuations.  

\section{Data of the Perseus cluster}
\label{sec:data}

We consider the data of the Perseus cluster from the {\it Gaia} DR2 catalog \citep{2016Gaia,2018Gaia} which is publicly available on the EAS Gaia archive\footnote{https://gea.esac.esa.int/archive/}. 
We consider the 32,672 stars brighter than $G=18$ mag within the circular region of radius 48 arcmins centered on the Perseus cluster: $\alpha$=2h~20m~45.3s (35.1889 deg), $\delta=50^\circ 7^\prime 50.5''$ (57.1307 deg). 
Our sample excludes the stars in the outer halo of Perseus, which is more extended than its central region, as discussed in \citet{2019Zhong}.

\begin{figure}[htp]
\includegraphics[width=0.45\textwidth]{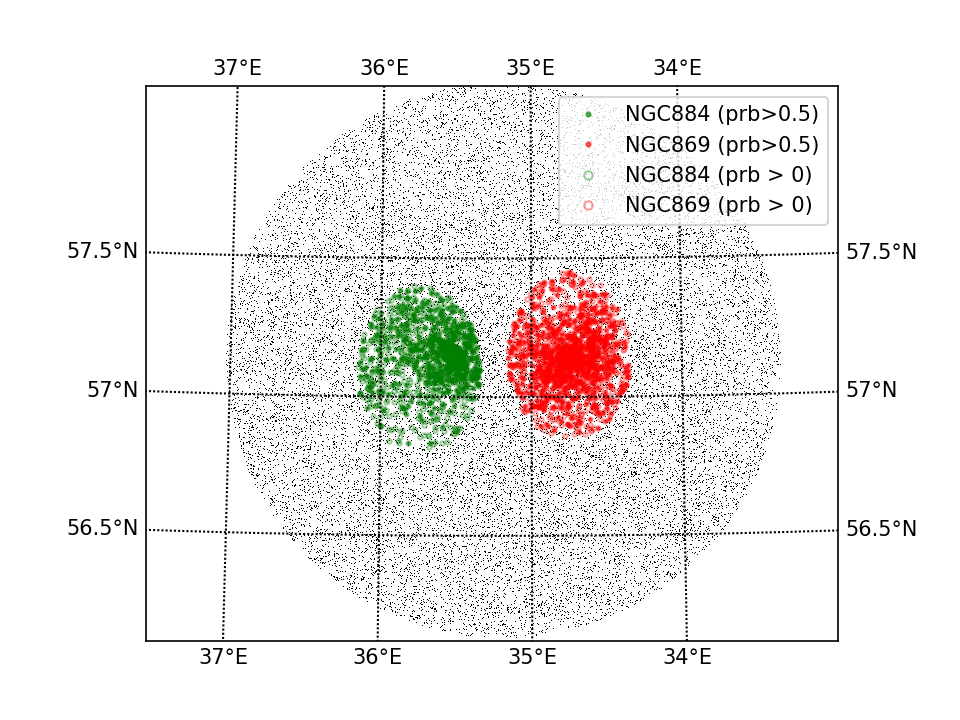}
	\caption{Distribution of the members of the Perseus double cluster according to \citet{GAIA2} in the azimuthal
	 equidistant projection. The grey dots show our sample of 32,672 stars brighter than $G=18$ mag.
	\citet{GAIA2} assign a membership probability ${\cal P}$ to the stars within two circular regions associated to NGC869 and NGC884 with predetermined centers and radii.
 The stars with membership probability ${\cal P}>0$ are shown by the colored symbols. The colored solid symbols show the stars with membership probability ${\cal P}>0.5$.} 
\label{fig:mem}
\end{figure}

We compare our analysis with the most recent identification of the members of Perseus performed by \citet{GAIA2}. 
They adopt the UPMASK method \citep{2014Krone} and use the celestial coordinates, proper motions and parallaxes provided by the {\it Gaia} DR2 sample. 
\citet{GAIA2} set the centers and the radii of two circular regions, and to each star within these regions they assign  
a membership probability according to the uncertainties on the proper motion and parallax of the star. 
The two circular regions are shown in Fig. \ref{fig:mem}: they have a diameter of 18 arcmins, 
and they do not overlap. The colored symbols in Fig. \ref{fig:mem}
show the members with membership probability ${\cal P}>0.5$. We also consider the stars with membership probability ${\cal P}>0$. 
The basic properties of these star samples are listed in Table \ref{table:sample}.

\begin{table*}[htb]
\caption{The basic properties of the Perseus double cluster according to \citet{GAIA2}}

\centering
\begin{tabular}{|c|c|c|c|c|c|c|}
\hline
\multirow{2}*{Name} & FoV\footnote{ Diameter of the field of view (FoV).}  & {$N_0$}\footnote{Number of stars brighter than $G=18$ mag in the FoV. For NGC869 and NGC884, it is the number of stars  with membership probability ${\cal P}>0$.} & $N_{0.5}$\footnote{Number of stars brighter than $G=18$ mag in the FoV with membership probability ${\cal P}>0.5$.} & D\footnote{Distance from the Sun according to \citet{2010Currie}. }  & $\mu_{RA}$\footnote{Mean proper motion with one standard deviation of the stars with membership probability ${\cal P}>0.5$.} & $\mu_{DEC}$ \\
           &  {\footnotesize (mag)} & ${\cal P}>0$  &  ${\cal P}>0.5$      & (pc)  &  (mas/yr) &  (mas/yr) \\
\hline
Field & 1.6    & 32672  &  -  &  - & - & - \\
NGC869 &  0.3    & 1422 & 720 & 2344$^{+88}_{-85}$  & -0.685 $\pm$ 0.131 & -1.074 $\pm$ 0.146 \\ 
NGC884 &  0.3    & 1107 & 483 & 2290$^{+87}_{-82}$  &  -0.614 $\pm$ 0.133& -1.058 $\pm$ 0.134\\
\hline
\end{tabular}
\label{table:sample}
\end{table*}

\begin{figure}[htp]
\includegraphics[width=0.45\textwidth]{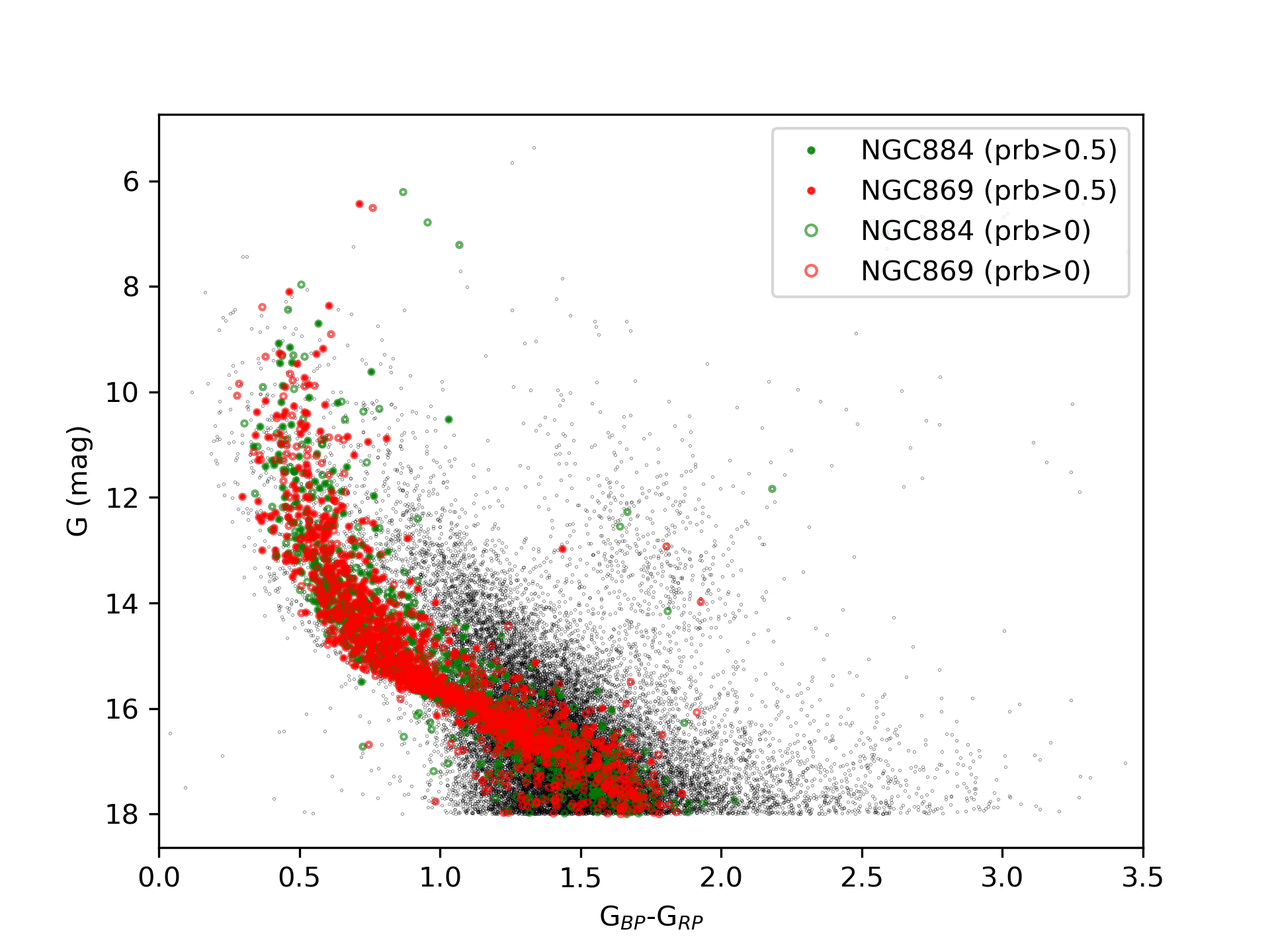}
\caption{Color-magnitude diagram of the members of the Perseus cluster according to the membership of \citet{GAIA2}. Colors and symbols are as in Fig. \ref{fig:mem}.
	The members of NGC884 and NGC869 have indistinguishable sequences.}
\label{fig:CMD}
\end{figure}

 Figure \ref{fig:CMD} shows the color-magnitude diagram of the system, from the accurate photometric data of the {\it Gaia} mission, based on the broad band $G$ magnitude, 
the blue $ G_{BP}$ (330 - 680 nm) and red $G_{RP}$ (640 - 1000 nm) colors.

\begin{figure}[htp]
\includegraphics[width=0.45\textwidth]{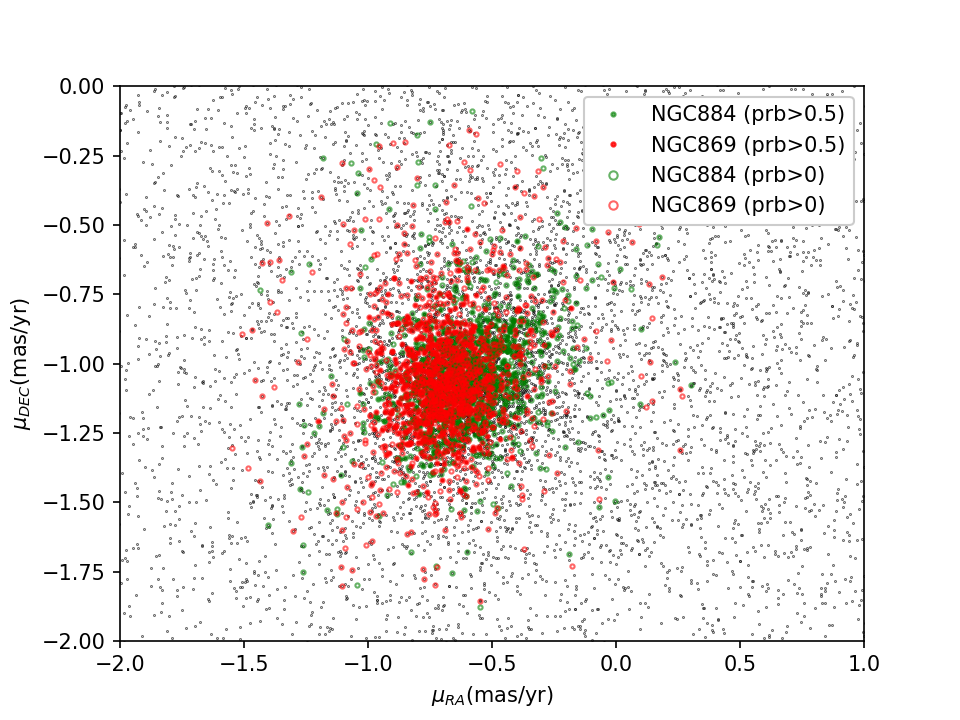}
	\caption{Proper motions of the Perseus members according to the membership of \citet{GAIA2}. Colors and symbols are as 
	in Fig. \ref{fig:mem}.}
\label{fig:pm}
\end{figure}

Figure \ref{fig:pm} shows the Perseus members in the plane of the components of the star proper motions. 
The distributions of these proper motions have mean and standard deviations in the two directions:
$\mu_{RA}=(-0.687\pm 0.217)$~mas yr$^{-1}$, $\mu_{DEC}=(-1.045\pm 0.239)$~mas yr$^{-1}$ for NGC869, 
and $\mu_{RA}=(-0.589\pm 0.217)$~mas yr$^{-1}$, $\mu_{DEC}=(-1.036\pm 0.220)$~mas yr$^{-1}$ for NGC884.

The standard deviations of these proper motion distributions coincide with the 
typical uncertainty on the proper motion $\sim 0.2$~mas~yr$^{-1}$ of $G = 17$~mag stars of the {\it Gaia} DR2 sample, which, at the distance of the Perseus cluster, 
corresponds to an uncertainty  of $\sim 2.2$~km~s$^{-1}$ on the velocity. 
These standard deviations are slightly larger than the estimated velocity dispersion $\sim 1.5$~km~s$^{-1}$ of the Perseus double cluster  \citep{2005Bragg}.
However, if we only consider stars with membership probability ${\cal P}>0.5$, the standard deviations reduce by almost a factor of $\sim 2$: 
the mean and standard deviation of the proper motions become 
$\mu_{RA}=(-0.685\pm 0.131)$~mas yr$^{-1}$, $\mu_{DEC}=(-1.074\pm 0.146)$~mas yr$^{-1}$ for NGC869, 
and $\mu_{RA}=(-0.614\pm 0.133)$~mas yr$^{-1}$, $\mu_{DEC}=(-1.058\pm 0.134)$~mas yr$^{-1}$ for NGC884, as listed in Table \ref{table:sample}.

\begin{figure}[htp]
 \centering
\includegraphics[width=0.45\textwidth]{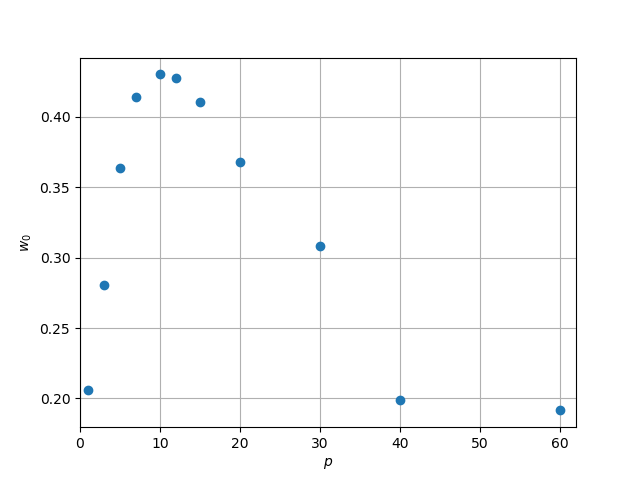}
\caption{The weight $w_0$ of the principal Gaussian component of the multi-Gaussian fitting procedure as a function of $p$. The largest $w_0$ corresponds to $p=10$. In addition to the values probed by the procedure described in Sect. \ref{sec:pvalue}, the figure also shows $w_0$ for $p=3$ and $p=60$.}
\label{fig:pmag18}
\end{figure}

\begin{figure*}[htp]
\includegraphics[width=0.95\textwidth]{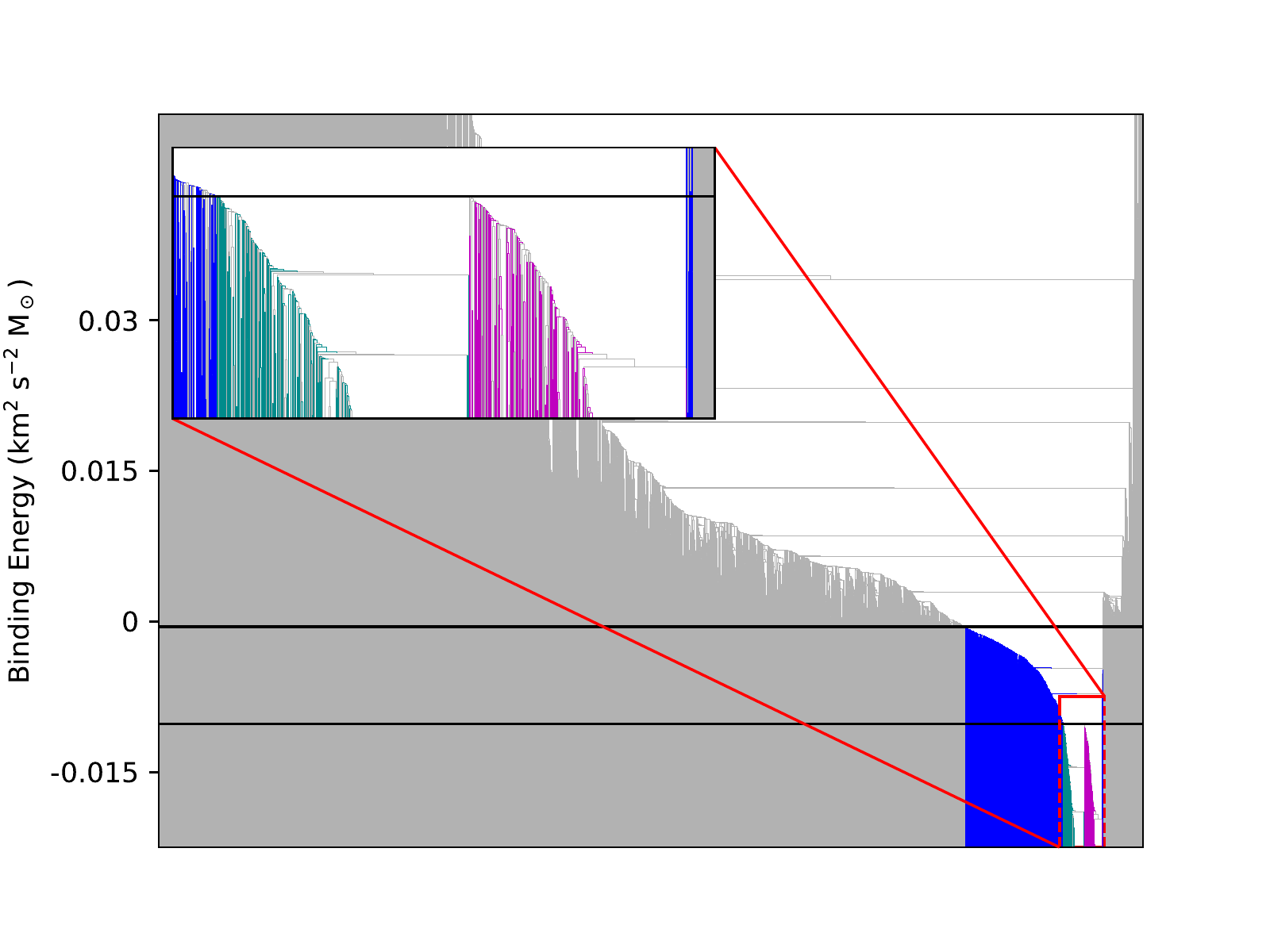} \\
\caption{Dendrogram of the Perseus double cluster with all the 32,672 stars in the field of view. The horizontal black lines are the two trimming thresholds. 
	The colored branches are the structures identified by the thresholds. The color code of the two structures is kept the same throughout the paper.}
\label{fig:tree}
\end{figure*}

\begin{figure}[htp]
\includegraphics[width=0.45\textwidth]{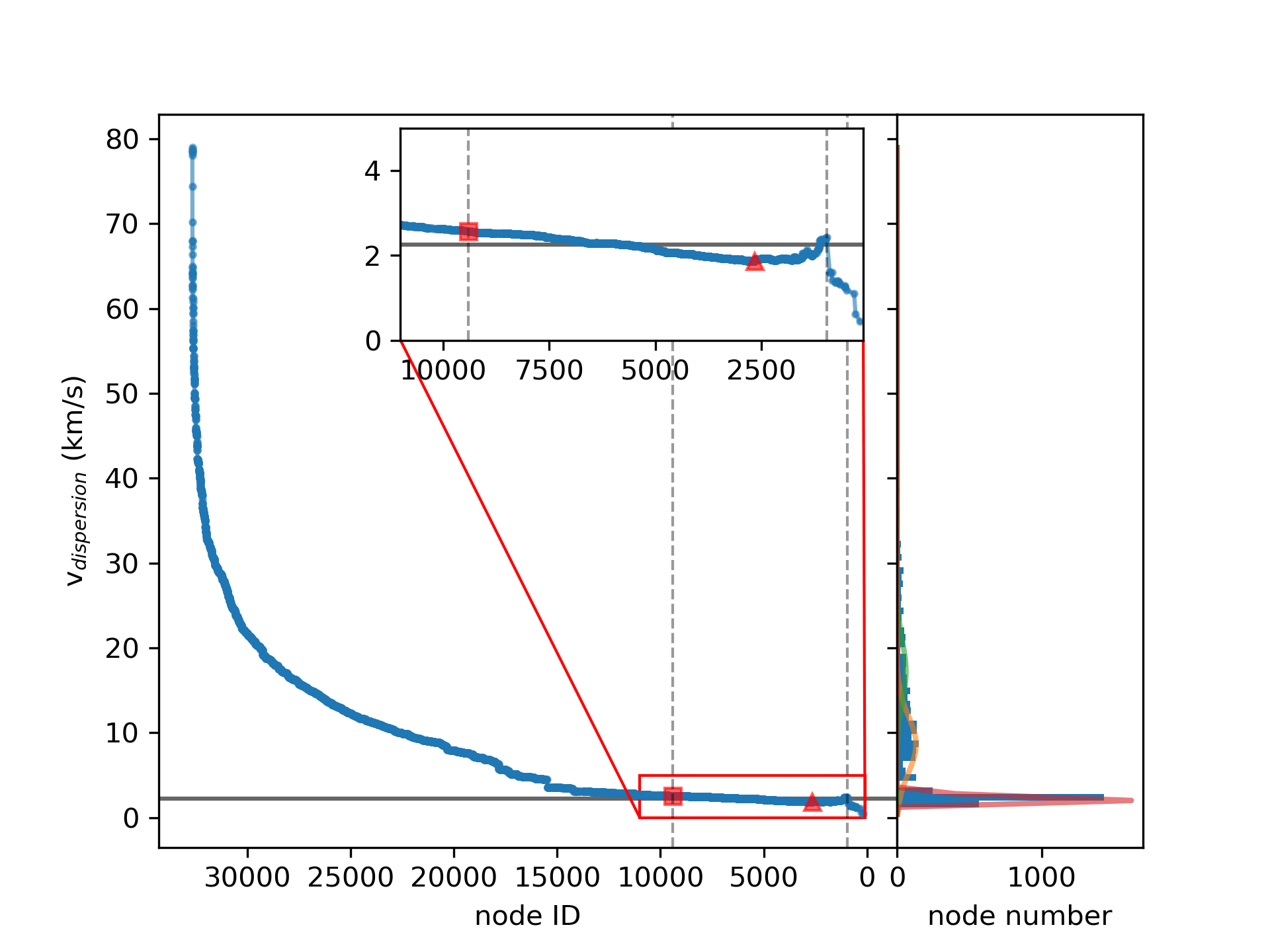} \\
\caption{ Velocity dispersions of the leaves hanging from the
nodes on the main branch of the dendrogram shown in Fig. \ref{fig:tree} as
a function of the node identification number. The root of the binary
tree is on the left; the leaves are on the right. The $\sigma$ plateau is shown by the  horizontal solid line, while its extension by the two vertical dashed lines. The red square and the red triangle are the two key nodes. The distribution of the velocity dispersions is on the right panel. Its multi-Gaussian fitting components are shown by the colored solid lines.}
\label{fig:mb}
\end{figure}

\section{The hierarchical structure of Perseus}
\label{sec:result}
We apply the method described in Sect. \ref{sec:method} to the positions and proper motions of the set of stars in the Perseus field of view described in the previous section. Figure \ref{fig:pmag18} shows that the algorithm sets to $p=10$ the optimal value of the parameter $p$ that identifies the $\sigma$ plateau. 
The hierarchical clustering method builds, according to the proxy of the pairwise binding energy, 
the binary tree shown in Fig. \ref{fig:tree}.

The velocity dispersions of the nodes on the main branch are shown in Fig. \ref{fig:mb}. The $\sigma$ plateau is at 2.26 $\pm$ 0.30 km s$^{-1}$; its location and extension are shown by the horizontal solid line and the two vertical dashed lines. The two key nodes of the $\sigma$ plateau, indicated by the red symbols, correspond to the thresholds at binding energies  $E=-0.0006$~km$^2$ s$^{-2}$ M$_\odot$ and $E=-0.0101$~km$^2$ s$^{-2}$ M$_\odot$, respectively. With these two thresholds, we identify the members of the Perseus cluster and its substructures.
\footnote{ The catalog of the members of Perseus and its substructures derived with our procedure is publicly available at
 \href{http://paperdata.china-vo.org/yuheng/paper/NGC0869\_mag18\_r0.8.mem.zip}{http://paperdata.china-vo.org/yuheng/paper/NGC0869\_mag18\_r0.8.mem.zip}. }

\subsection{The main cluster Sub1}
\label{sec:upperthreshold}

\begin{figure}[htp]
\includegraphics[width=0.49\textwidth]{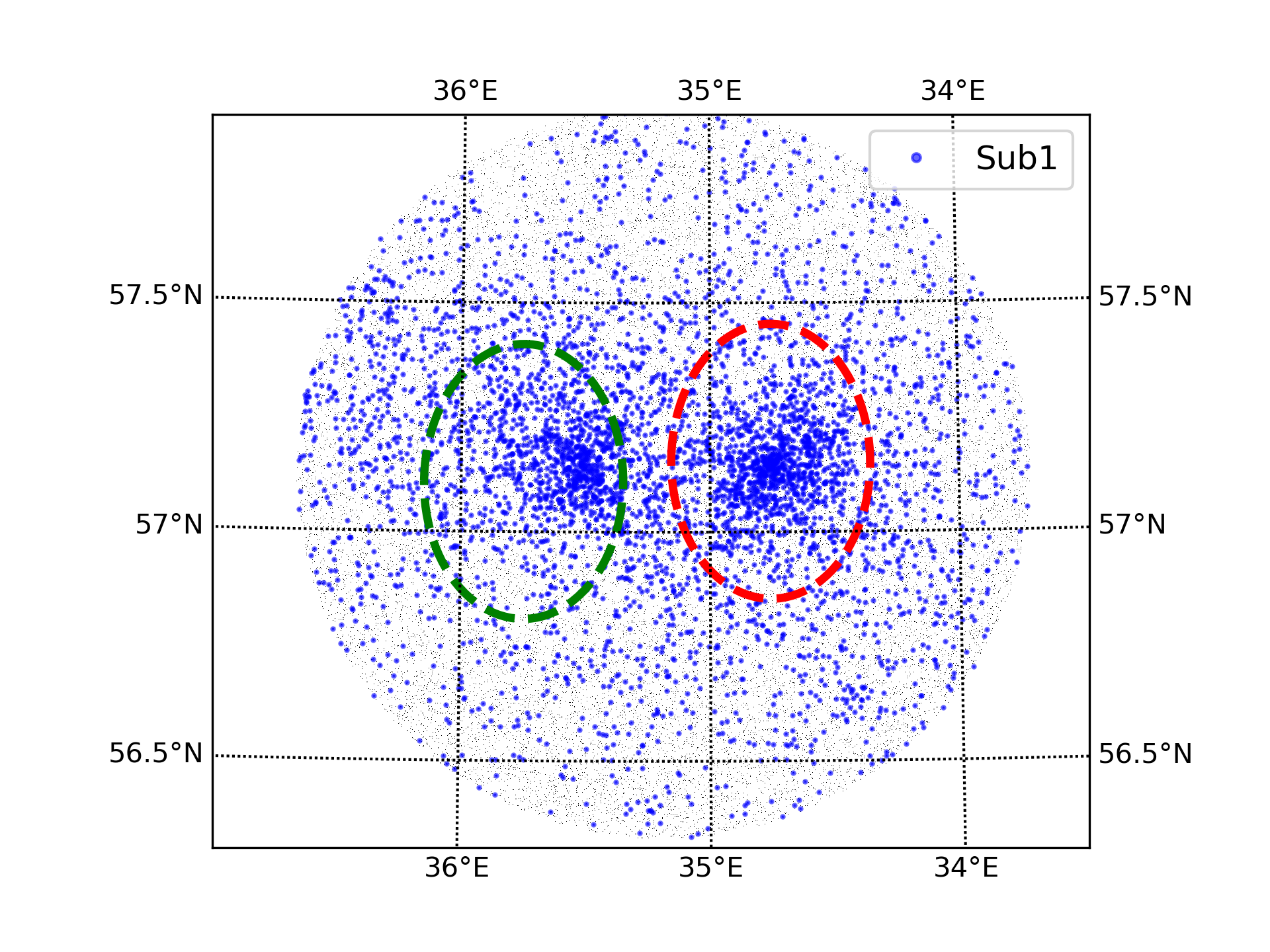}
	\caption{The blue dots show the distribution on the sky of the members of Sub1 identified by trimming the binary tree shown in Fig. \ref{fig:tree} with the threshold corresponding to the left key node of the $\sigma$ plateau shown in Fig. \ref{fig:mb}. The grey dots show the remaining stars in the field of view. The two circles indicate the regions set by \citet{GAIA2}.}
\label{fig:perseus1}
\end{figure}

With the upper threshold $E$=-0.0006 km$^2$ s$^{-2}$ M$_\odot$, the binary tree  returns only one structure with more than 
100 members: it contains 4542 members. 
We call this structure Sub1. Figure \ref{fig:perseus1} shows the distribution of its members on the sky. Its extended shape supports the conclusion
of \citet{2019Zhong} that the physical scale of the double cluster is much larger than the size of its core. 

\begin{figure}[htp]
\includegraphics[width=0.49\textwidth]{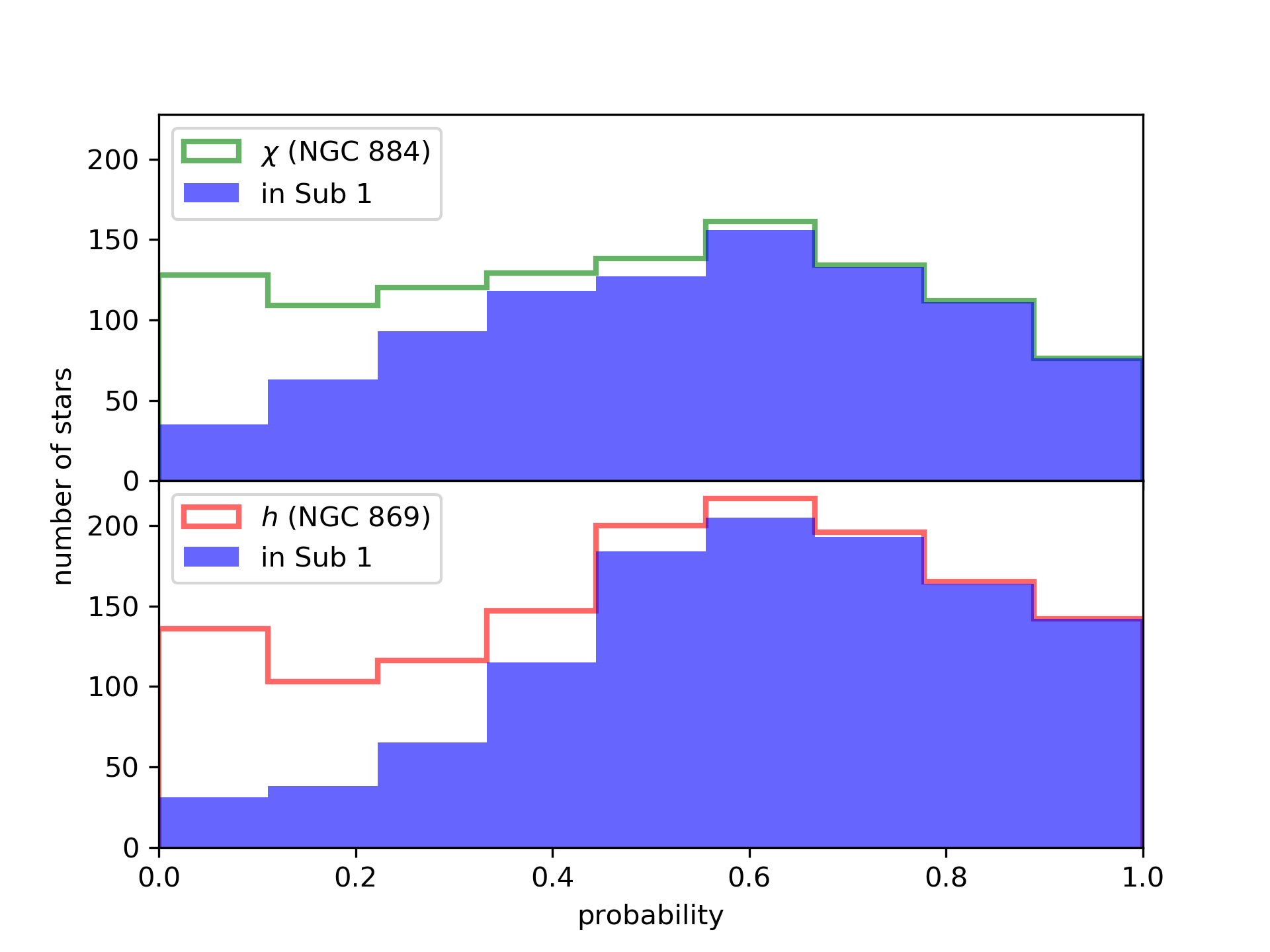}
	\caption{The open histograms show the distributions of the membership probability ${\cal P}$,  according to \citet{GAIA2}, of the stars with ${\cal P}>0$ in the circular regions of NGC884 or NGC869 shown in Fig. \ref{fig:mem}. The solid histograms show the distributions of ${\cal P}$ of  the subset of these stars that are also members of Sub1. 
	The upper and lower panel shows the distributions for NGC884 and NGC869, respectively.
}
\label{fig:ratio1-2}
\end{figure}

\begin{figure}[htp]
\includegraphics[width=0.49\textwidth]{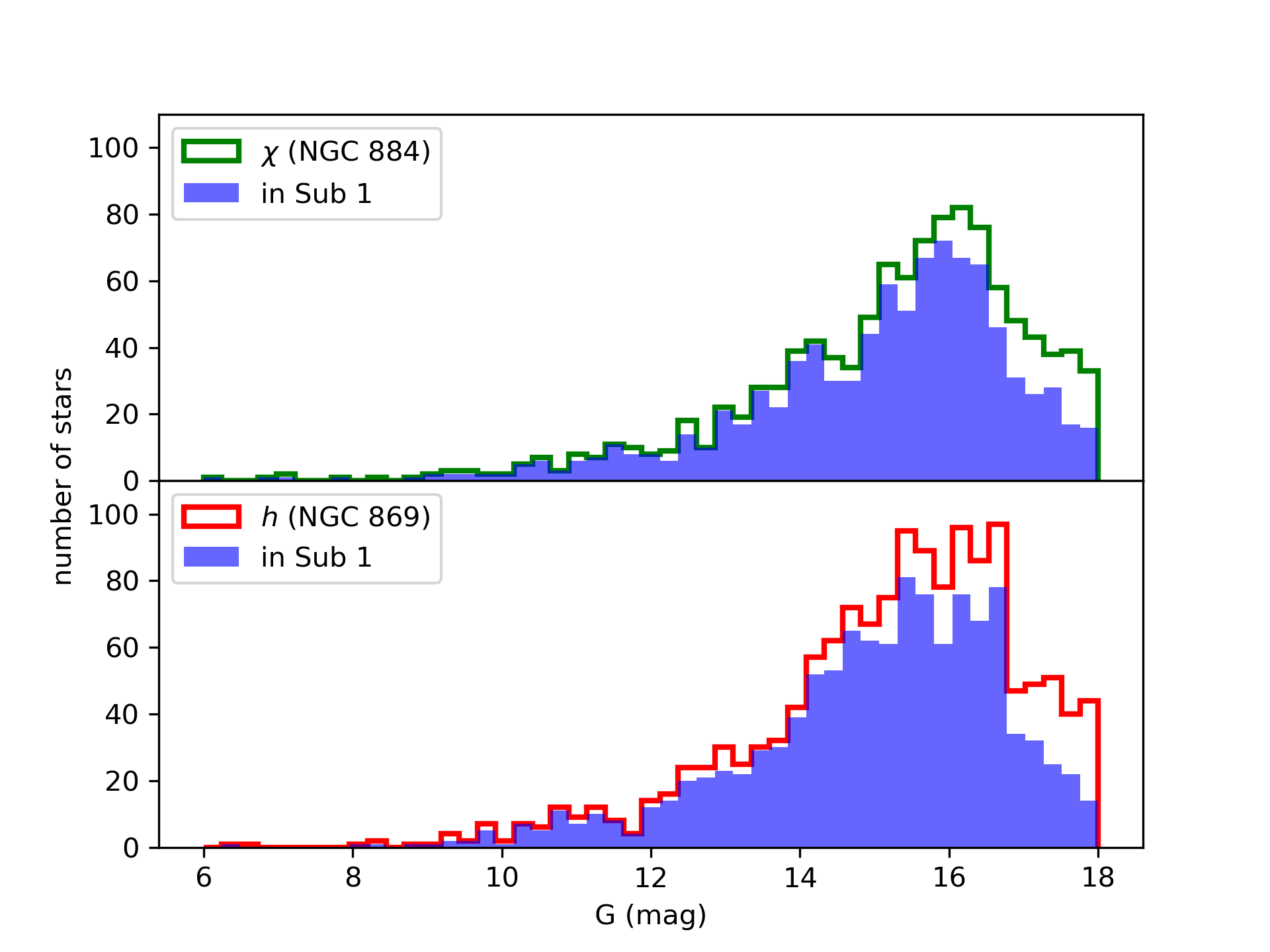} \\
	\caption{ Same as Fig. \ref{fig:ratio1-2} for the distribution of the star magnitudes. 
}
\label{fig:ratio1-1}
\end{figure}

We compare the members of Sub1 with the members of NGC869 and NGC884 identified by \citet{GAIA2}: 
1137 (912) members of Sub1 are stars of NGC869 (NGC884), whose membership probability ${\cal P}$ computed by \citet{GAIA2} is larger than $0$. Figure \ref{fig:ratio1-2} shows the distributions of ${\cal P}$ of these stars. The Sub1 members tend to have large ${\cal P}$, whereas a large fraction of stars that \citet{GAIA2} associate to small ${\cal P}$ are not identified as members of Sub1. Specifically,  
for NGC869, Sub1 contains 97.8\% of the \citet{GAIA2} stars with ${\cal P}>0.5$, namely 704 stars out of 720; similarly, for NGC884, Sub1 contains 98.6\% of the ${\cal P}>0.5$ stars, namely 476 stars out of 483 (see also Table \ref{table:compare}). When considering the stars with ${\cal P}\le 0.5$, Sub1 contains 61.7\% (433 stars out of 702) and  
69.9\% (436 stars out of 624) of the \citet{GAIA2} stars, for NGC869 and NGC884, respectively.

Figure \ref{fig:ratio1-1} shows the distributions of the magnitudes of the stars with ${\cal P}>0$, according to \citet{GAIA2}, in the circular regions of NGC884 or NGC869, and the distributions of the  magnitudes of the subsets of these stars that are also members of Sub1.
We recover most stars of \citet{GAIA2}  with ${\cal P}>0$ brighter than $G=16$ mag: Sub1 includes 782 out of the 902 (86.7\%) bright members of NGC869 and 603 out of the 674 (89.5\%) bright members of NGC884 of \citet{GAIA2}.

\begin{figure}[htp]
\includegraphics[width=0.49\textwidth]{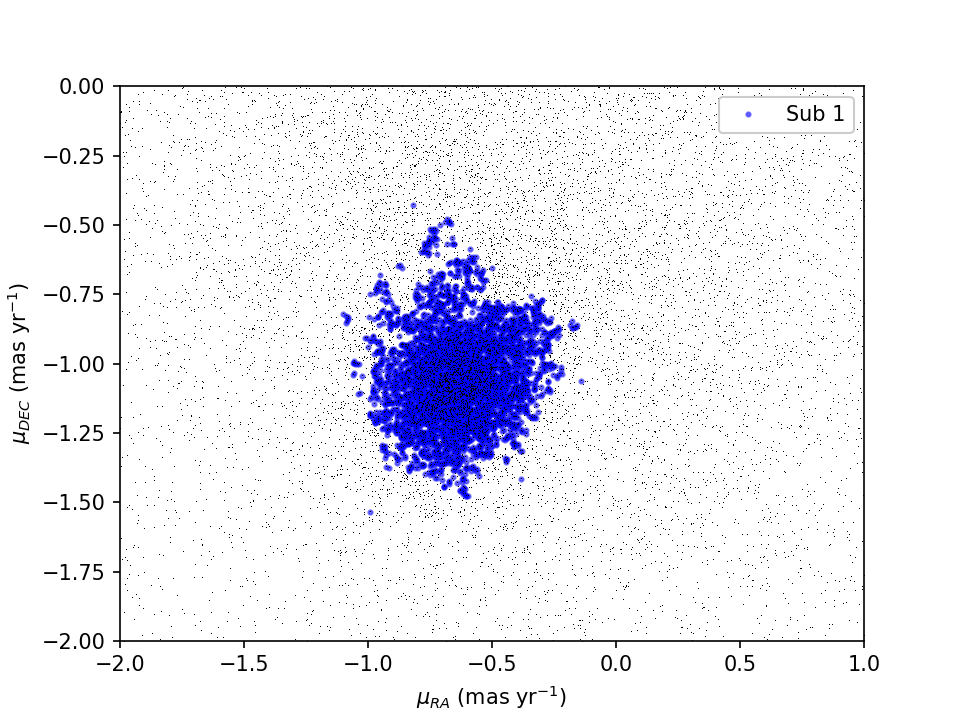}
	\caption{The blue dots show the distribution of the proper motions of the  members of Sub1 in the plane of the proper motion components. The grey dots show the remaining stars in the field of view.
The axis ranges are the same as in Fig. \ref{fig:pm}.}
\label{fig:perseus1_pm}
\end{figure}

Figure \ref{fig:perseus1_pm} shows the distribution of the members of Sub1 in the plane of the components of the star proper motions.
The Sub1 members are concentrated within a roughly circular region of radius $\sim 0.5$~mas~yr$^{-1}$. 
The distribution has mean and standard deviation in the two directions: $ \mu_{\mathrm {RA}} =(-0.640 \pm 0.160) $ mas~yr$^{-1}$ and  $ \mu_{\mathrm {DEC}} = (-1.048 \pm 0.165)$ mas~yr$^{-1}$.
These standard deviations are in between the standard deviations $\sim 0.22-0.24$~mas~yr$^{-1}$ of NGC884 and NGC869, estimated with the stars 
 of \citet{GAIA2}  with ${\cal P}>0$, and the standard deviations $\sim 0.12-0.14$~mas~yr$^{-1}$, 
 estimated with the stars with ${\cal P}>0.5$, as reported in Sect. \ref{sec:data}.

\begin{figure}[htp]
\includegraphics[width=0.49\textwidth]{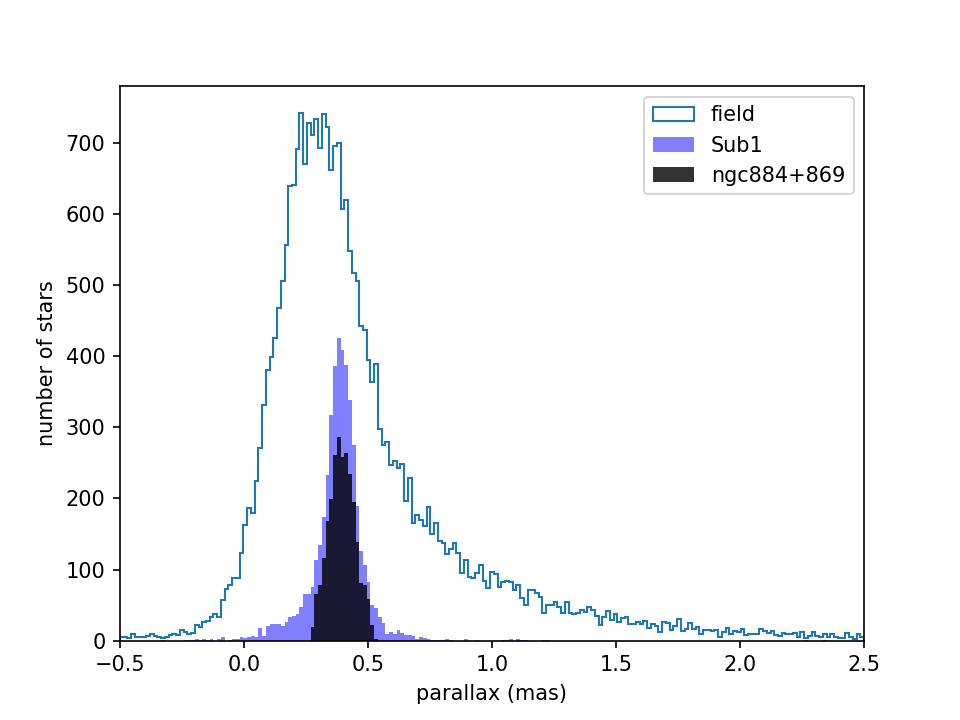}
	\caption{ Distributions of the star parallaxes. The blue solid histogram is the distribution of the members of Sub1. The black solid histogram is the
combined distribution of the stars of NGC869 and NGC884 with membership probability ${\cal P}>0$ according to \citet{GAIA2}. The blue open histogram is the distribution of the stars in the field of view that are not members of Sub1.  }
\label{fig:para}
\end{figure}

Figure \ref{fig:para} shows the  distribution of the parallaxes of the Sub1 members. Similarly to \citet{GAIA2}, we added the zero-point offset 0.029 mas to each star parallax to correct for the systematic errors of the {\it Gaia} DR2\footnote{ According to \citet{2019Zinn}, this offset derives from the degeneracy in the astrometric solution between the global parallax shift and the term describing
the periodic variation of the spacecraft's basic angle with the spacecraft spin period.} . The mean and
standard deviations $\varpi = 0.406 \pm 0.074$~mas
match the values of the two clusters based on the stars with membership probability 
 ${\cal P}>0.5$ of \citet{GAIA2}: $\varpi = 0.400 \pm 0.042$~mas for NGC869  
and $\varpi = 0.398 \pm 0.039$~mas for NGC 884.

\begin{figure}[htp]
\includegraphics[width=0.49\textwidth]{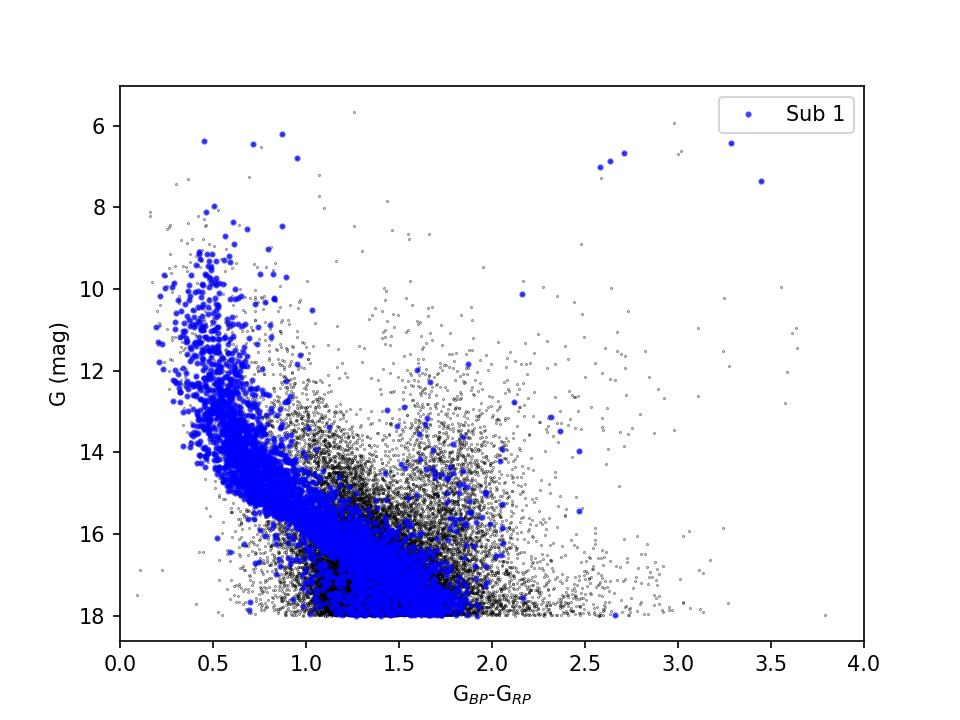} 
	\caption{ The blue dots show the distribution of the  members of Sub1 in the color-magnitude diagram. The grey dots show the remaining stars in the field of view.   
The axis ranges are the same as in Fig. \ref{fig:CMD}.}
\label{fig:perseus1_cmd}
\end{figure}

Finally, Fig. \ref{fig:perseus1_cmd} shows that the color-magnitude diagram of the Sub1 members is 
qualitatively comparable with the color-magnitude diagrams of the \citet{GAIA2}  stars with membership probability ${\cal P}>0$ of NGC884 and NGC869  shown in Fig. \ref{fig:CMD}.

Overall, Figures \ref{fig:tree}-\ref{fig:perseus1_cmd} show that the main cluster Sub1 identified by our hierarchical algorithm contains the two
components NGC884 and NGC869. They are two substructures of the larger Perseus cluster, whose gravitational hierarchy is illustrated in the dendrogram of Fig. \ref{fig:tree}. In the next subsection, we show that, by  adopting the lower threshold to trim the binary tree, our
algorithm is able to separate Sub1 into the two expected substructures.

\subsection{The substructures Sub1-1 and Sub1-2}

A relevant advantage of the hierarchical clustering method is that, 
once the stars are arranged in the binary tree, different structures in the field of view 
can be immediately identified by the proper trimming threshold.

Figure \ref{fig:tree} shows that the binary tree splits into two separate structures at the binding energy $E = -0.0101$ km$^2$ s$^{-2}$ M$_\odot$,  
shown by the lower horizontal line. 
This lower threshold is set by the rightmost key node of the $\sigma$ plateau shown in Fig. \ref{fig:mb}.
Adopting this threshold removes
most of the members of Sub1 in the cluster outskirts and 
focuses on the deepest region of the gravitational potential well of the Perseus double cluster. 

The two substructures with more than 100 members identified by this  lower threshold, Sub1-1 and Sub1-2, 
are substructures of Sub1.
Their basic properties are listed in Table \ref{table:sub}. We also list the properties of 
Sub1-0, the system of stars that are members of Sub1, but are not members of either Sub1-1 or Sub1-2.

\begin{table}[htb]
\caption{Properties of the Perseus double cluster according to the binary tree structure}
\centering
\begin{tabular}{|c|c|c|c|}
\hline
	ID & N$_{\mathrm {mem}}$\footnote{ Number of the members of the binary tree structures.} &   $\mu_{\mathrm {RA}}$\footnote{ Mean and standard deviation of the proper motions of the members of the binary tree structures.} (mas/yr) & $\mu_{\mathrm {DEC}}$ (mas/yr) \\
\hline
Sub1  &  4542 &       $-0.640 \pm 0.160$ & $-1.048 \pm 0.165$  \\
\hline
Sub1-1 &  695  &   $ -0.686 \pm 0.101 $ & $-1.074 \pm 0.139 $ \\
Sub1-2 &  598  &   $ -0.614 \pm 0.110 $ & $-1.074 \pm 0.111 $ \\
Sub1-0 &  3249  &   $ -0.629 \pm 0.176 $ & $-1.046 \pm 0.178 $\\
\hline
\end{tabular}
\label{table:sub}
\end{table}

\begin{table}[htb]
\caption{ Membership probability of the \citet{GAIA2} members that are also members
of the binary tree structures}
\centering
\begin{tabular}{|c|cc|cc|}
\hline
  & \multicolumn{2}{c|}{NGC869} & \multicolumn{2}{c|}{NGC884} \\
  & ${\cal P}>0$ & ${\cal P}>0.5$ & ${\cal P}>0$ & ${\cal P}>0.5$ \\
\hline
UPMASK\footnote{ Number of members according to \citet{GAIA2}.} &  1422 & 720  & 1107 & 483 \\
\hline
Sub1-1 &  622 & 444 & 0 & 0 \\
Sub1-2 & 17 & 13 & 444 & 299 \\
Sub1-0 & 498 & 247 & 468 & 177 \\
Sub1 & 1137 & 704 & 912 & 476 \\
\hline
\end{tabular}
\label{table:compare}
\end{table}

\begin{figure}[htp]
\includegraphics[width=0.49\textwidth]{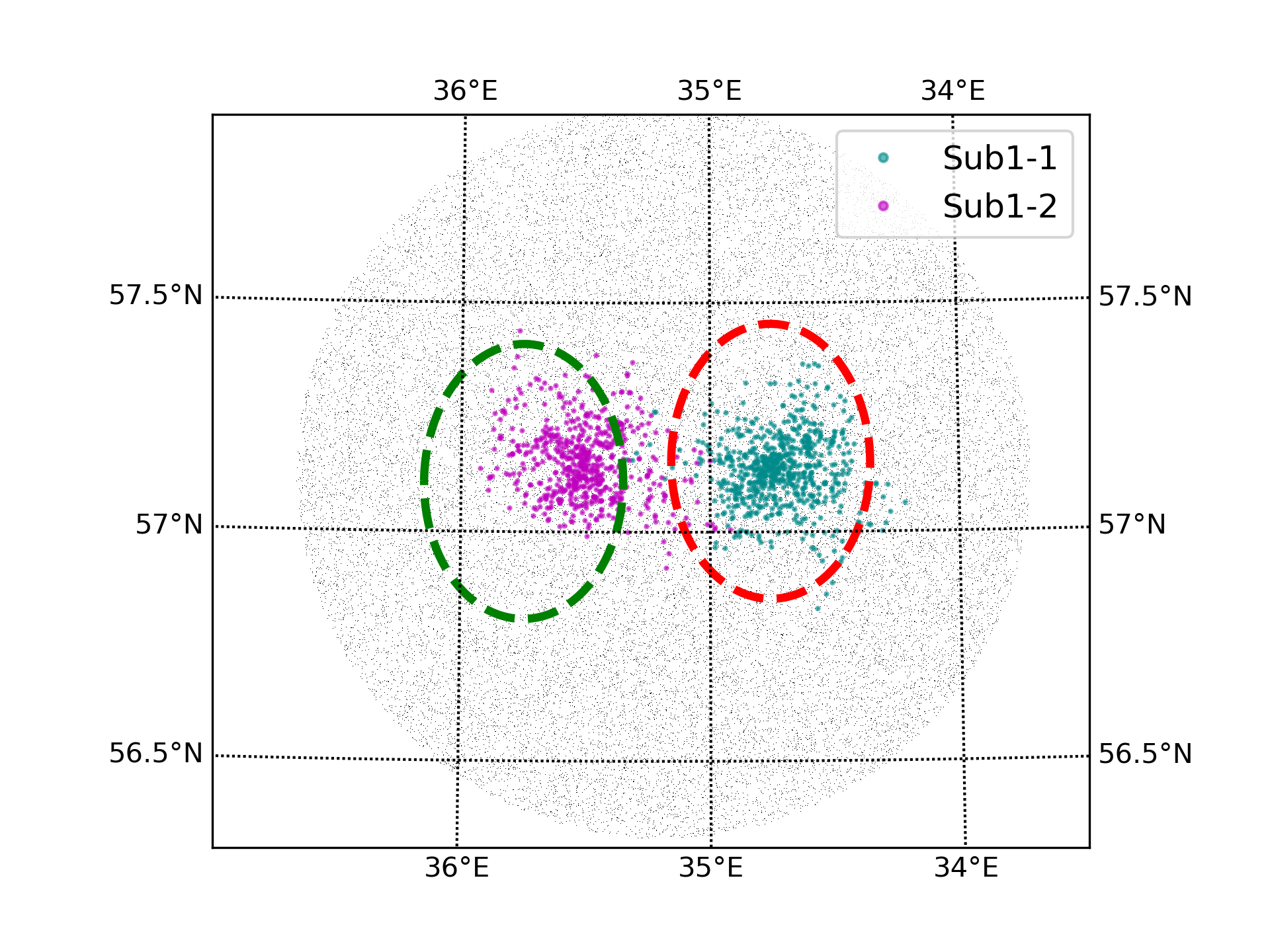}
	\caption{ Distribution of the members  of Sub1-1 (cyan dots, mostly on the right) and Sub1-2 (magenta dots, mostly on the left) on the sky. 
The grey dots show the entire star sample. The two circles indicate the regions set by \citet{GAIA2}.}
\label{fig:perseus2}
\end{figure}

Figure \ref{fig:perseus2} shows the distribution of the members of Sub1-1 and Sub1-2 on the sky.
Sub1-1 and Sub1-2 overlap with NGC869 and NGC884, respectively.
Compared to NGC869 and NGC884, Sub1-1 and Sub1-2 contain fewer members  than the stars
with membership probability ${\cal P}>0$ identified by \citet{GAIA2}, as listed in Tables \ref{table:sub} and \ref{table:compare}: 695 and 598, compared with 1422 and 1107 for NGC869 and NGC884, respectively. However,  89.5\% of the Sub1-1 members, 622 out of 695 stars, are \citet{GAIA2} members of NGC869 with ${\cal P}>0$. For Sub1-2 and NGC884, this fraction is 74.2\%, namely 444 out of 598 stars. 

Similarly to Sub1, Sub1-1 and Sub1-2 tend to have members with large ${\cal P}$: Table \ref{table:compare} shows that 71.4\% of the ${\cal P}>0$ members of NGC869 have ${\cal P}>0.5$, namely 444 out of 622 stars; this percentage is 67.3\% for NGC884, namely 299 out of 444 stars.  
Table \ref{table:compare} shows that Sub1-2, that overlaps with NGC884, also contains 17 stars of the ${\cal P}>0$ stars
of NGC869, suggesting that the evident separation between NGC869 and NGC884 on the sky might not be fully complete.

\begin{figure}[htp]
\includegraphics[width=0.49\textwidth]{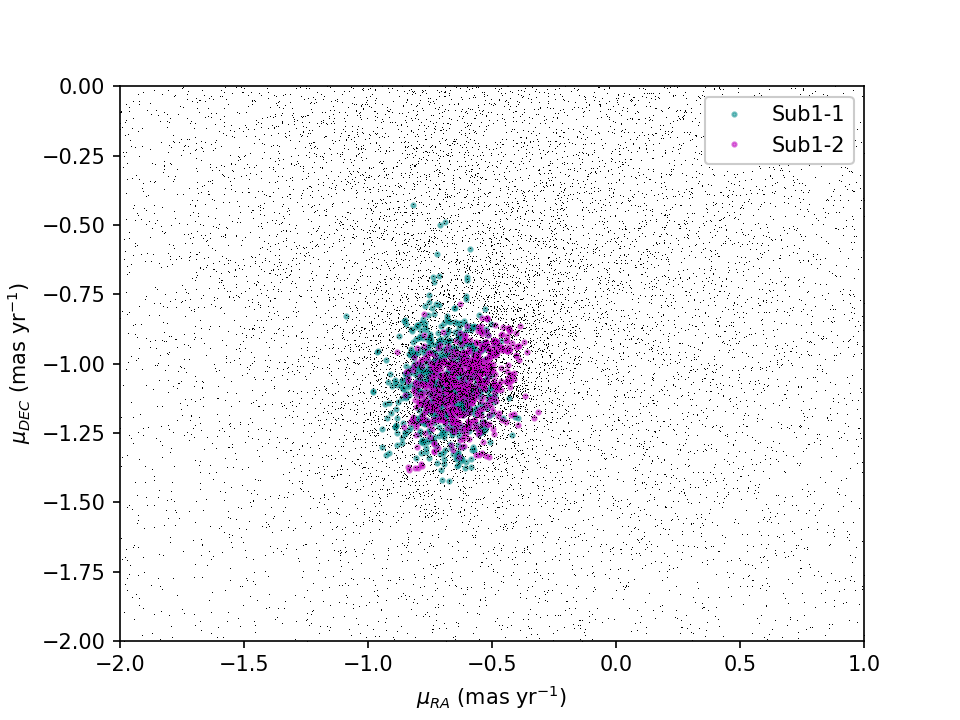}
	\caption{Distributions of the proper motions of the members of Sub1-1 and Sub1-2. Colors are as in Fig. \ref{fig:perseus2}.}
\label{fig:perseus2_pm}
\end{figure}

Figure \ref{fig:perseus2_pm} shows that the distributions of the components of the proper motions of the members
of Sub1-1 and Sub1-2 are more concentrated
than the distributions of the stars  of \citet{GAIA2} with  ${\cal P}>0.5$ shown in Fig. \ref{fig:pm}.
According to Tables \ref{table:sample} and \ref{table:sub}, Sub1-1 has velocity dispersion 
in the two directions $(\sigma_{\mathrm {RA}},\sigma_{\mathrm {DEC}})=(0.101,0.139)$~mas~yr$^{-1}$, 
whereas  NGC869 has 
$(\sigma_{\mathrm {RA}},\sigma_{\mathrm {DEC}})=(0.121, 0.133)$~mas~yr$^{-1}$. Similarly, 
Sub1-2 has $(\sigma_{\mathrm {RA}},\sigma_{\mathrm {DEC}})=( 0.110,0.111)$~mas~yr$^{-1}$, whereas NGC884 has 
$(\sigma_{\mathrm {RA}},\sigma_{\mathrm {DEC}})=(0.124,0.122)$~mas~yr$^{-1}$. Therefore, unlike the velocity dispersion of Sub1 reported
in Sect. \ref{sec:upperthreshold}, the velocity dispersions of Sub1-1 and Sub1-2 are smaller than the velocity dispersions of  NGC869 and NGC884, indicating
that our hierarchical algorithm identifies members with more similar proper motions than
\citet{GAIA2}.

\begin{figure}
\includegraphics[width=0.49\textwidth]{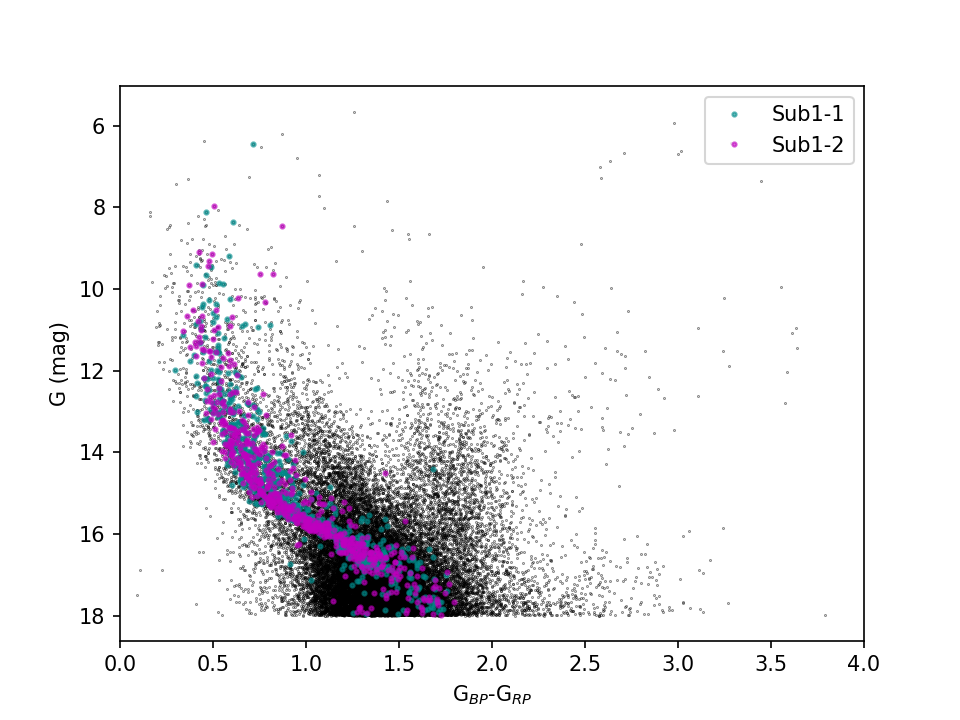} \\
\caption{Distributions of the members of Sub1-1 and Sub1-2 in the color-magnitude diagram. Colors are as in Fig. \ref{fig:perseus2}.}
\label{fig:perseus2_cmd}
\end{figure}

Figure  \ref{fig:perseus2_cmd} shows the color-magnitude relations of the members of Sub1-1 and Sub1-2: they are similar to each other and qualitatively similar to the color-magnitude relations of NGC869 and NGC884 shown in Fig. \ref{fig:CMD}. These similarities  support the conclusion of \citet{2002Slesnick} that NGC869 and NGC884, and thus Sub1-1 and Sub1-2, have  the same epoch of star formation. 
The scatter along the color-magnitude ridge line of Sub1-1 is $\sim 0.183$ mag and is comparable to the scatter $\sim 0.184$ mag of 
the stars of NGC869 with membership probability ${\cal P}>0.5$. For NGC884, 
the stars with ${\cal P}>0.5$  have scatter $\sim 0.179$ mag, which is $\sim 15$\%  larger than the scatter of Sub1-2 $\sim 0.155$ mag. 
This result shows that our algorithm 
identifies members with more similar photometric properties than the approach of \citet{GAIA2}.
 
All these results indicate that Sub1-1 and Sub1-2 coincide with the traditional clusters NGC869 and NGC884. Our results also confirm that the approach of \citet{GAIA2}, who, unlike our algorithm, predetermined the centers and sizes of  NGC869 and NGC884, is legitimate for Perseus, because, although 
Sub1-1 and Sub1-2 are very close to each other, they remain largely distinct on the sky.

\begin{figure}[htp]
\includegraphics[width=0.49\textwidth]{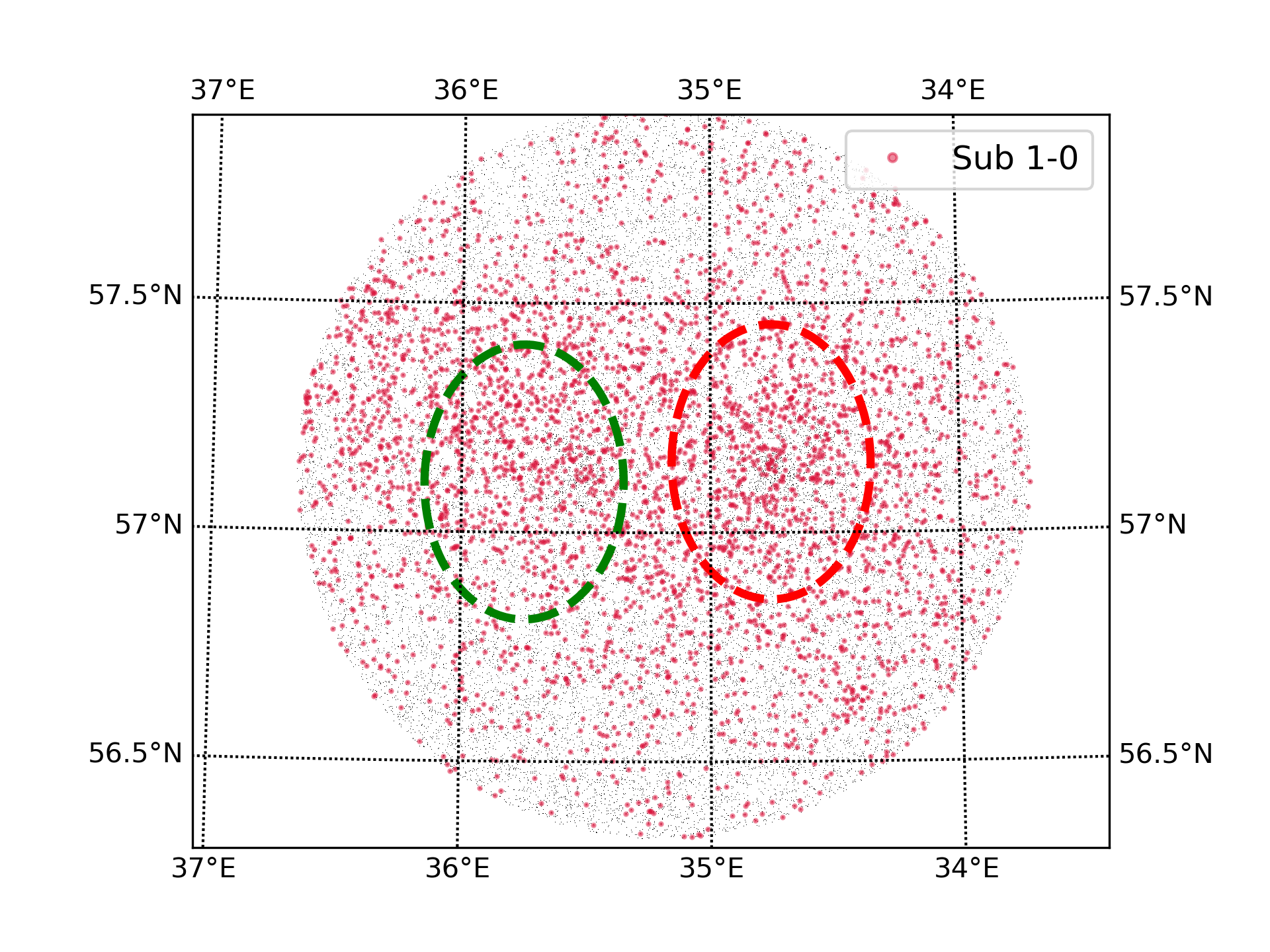}
	\caption{ Distribution of the halo members of the Perseus double cluster on the sky. 
	The grey dots show our entire star sample. The crimson symbols
	show the members of the cluster halo Sub1-0.
	The two circles indicate the regions set by \citet{GAIA2}.}
\label{fig:halo}
\end{figure}

The star members of Sub1 which do not belong to either Sub 1-1 or Sub 1-2 are stars of the halo of the cluster. We label this halo as Sub1-0. The distribution of these stars on the sky, shown in Fig. \ref{fig:halo}, shows that they are almost uniformly distributed over the field.

\begin{figure}
\includegraphics[width=0.49\textwidth]{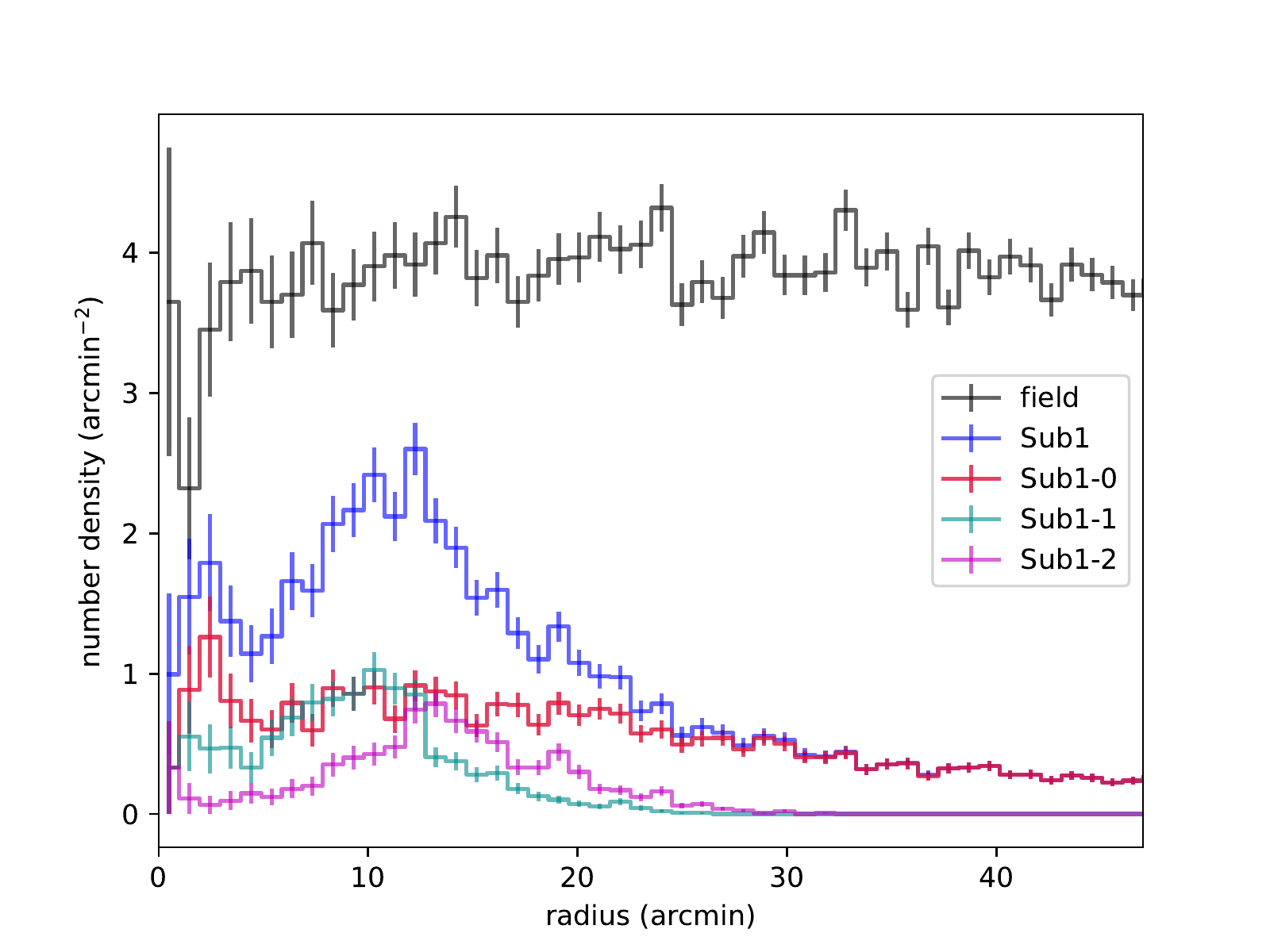} \\
\caption{ Radial profiles of the star number density for the different structures identified by the binary tree. The number density is estimated in 
circular rings centered on the center of the field $(\alpha,\delta)=(35.1889,57.1307)$~deg.
The black  profile is for the field stars, namely stars that are not members of Sub1.
The blue profile is for the Sub1 members. The red profile is for the stars of Sub1-0, while the cyan and magenta profiles are for Sub1-1 and Sub1-2, respectively. The errors are $1-\sigma$ standard deviations of Poisson variates.}
\label{fig:radial}
\end{figure}

The difference between the spatial distribution of the members of the binary tree structures and the remaining stars in the field is further illustrated in Fig. \ref{fig:radial}, that shows the radial profiles of the star number densities of the different structures. 
The black line in the top is  the profile of the field stars, namely all the stars in our catalog that are not members of Sub1. This profile is roughly constant and
 it drops in the very center for a statistical fluctuation caused by the small area.
The centers of the two components Sub1-1 and Sub1-2 are at $\sim 12$ arcminutes from the center of Perseus, and determine the peak of the Sub1 profile shown in blue. The profiles of Sub1-1 and Sub1-2 , the cyan and magenta profiles respectively, display a similar peak in correspondence of the peak of Sub1.
The halo stars Sub 1-0, whose profile is in red, have a relatively flat distribution, similarly to the field stars whose profile is in black.
These halo stars account for the wide spread both in the proper motion diagram shown in Fig. \ref{fig:perseus1_pm} and in the color-magnitude diagram shown in Fig. \ref{fig:perseus1_cmd}.

\section{Conclusion }
\label{sec:end}

 We propose a hierarchical clustering algorithm to identify the members and the substructures of open star clusters. The algorithm requires at least the celestial coordinates and the proper motions of the stars in the field of view of the cluster. We test our algorithm on star mock catalogs and apply it to the Perseus cluster.

Our hierarchical clustering algorithm is based on the single-linkage method, where 
we adopt a proxy of the pairwise binding energy as the distance metric to arrange the stars in the
field of view in a binary tree. We use the $\sigma$-plateau method to trim the binary tree and
associate its branches to the main cluster and its substructures; the leaves of these branches are  the star members of the structures.
Our algorithm  relies neither on the photometric properties of the stars nor on any assumption on the shape, size, evolutionary or dynamical state of the cluster.
The algorithm is thus ideal  to investigate unrelaxed irregular systems like  open clusters.

When applied to the {\it Gaia} DR2 data in the field of view of the Perseus cluster, the proxy for the  pairwise binding energy associates the same mass $m=1$~M$_\odot$ to all the stars in the field of view and uses only four out of the six phase-space coordinates of each star: the celestial coordinates and the two components of the proper motion. We ignore the radial
distances and the radial velocities of the stars, because their uncertainties are much larger than the size
and the velocity dispersion of the cluster. 

We test the algorithm with this proxy of the binding energy on 100 mock catalogs mimicking the Perseus field of view. The algorithm correctly identifies the cluster: it returns a completeness of $(91.5\pm 3.5)$\% and a fraction of interlopers of $(10.4\pm 2.0)$\%; the same algorithm where the proxy for the binding energy includes all the six phase-space coordinates returns a completeness of $(96.7\pm 2.6)$\% and an interloper fraction of $(3.2\pm 2.0)$\%: the bias introduced by estimating the proxy for the binding energy with four coordinates alone is thus within the statistical fluctuations. 

The algorithm applied to the stars in the Perseus field of view identifies the cluster members and separates the cluster into two distinct 
substructures that are located in the deepest region
of the cluster gravitational potential well. 
These two substructures, Sub1-1 and Sub1-2,  correspond to NGC869 ($h$ Per) and NGC884 ($\chi$ Per),
respectively.
In fact, their members share the same photometric and kinematic properties of the members of NGC869  
and NGC884 identified within two regions on the sky set a priori by \citet{GAIA2}.
Compared to this latter analysis, the velocity dispersions 
of the members of Sub1-1 and Sub1-2 are 5\% to 23\% smaller, depending on the  proper motion
component. Similarly, the scatter around the color-magnitude relation is comparable for Sub1-1 and NGC869, whereas the scatter is $\sim 15\%$ smaller in Sub1-2 compared with NGC884. 

These results suggest that our
algorithm identifies members that have more homogeneous kinematic and photometric properties than the procedure adopted by \citet{GAIA2}, 
despite the fact that our algorithm is only based on the star proper motions and ignore the
photometric and spectroscopic properties of the stars.

Our hierarchical clustering algorithm is an efficient tool that can be easily applied to other data sets.  With the high-accuracy data coming from future spectroscopic and astrometric surveys \citep[e.g.][]{Theia2017, 2019Malbet}, that are expected to increase the accuracy reached by, e.g., SEGUE \citep{2009Yanny}, RAVE \citep{2013Kordopatis}, APOGEE \citep{2017Majewski}, Gaia \citep{2018Gaiadr2} and LAMOST \citep{2012Lamost}, the proxy for the binding energy can be improved by (i) including all the six phase-space coordinates and (ii) assigning the proper mass to each star. These enhancements will further increase the ability of our algorithm to unveil the complex inner structure of open clusters and understanding their formation and evolution.

\software{astropy \citep{2013A&A...558A..33A},  
          scikit-learn \citep{scikit-learn}
          }
          
\acknowledgments
We dedicate this article to the 60th anniversary of the Department of Astronomy of Beijing Normal University,
the 2nd astronomy programme in the modern history of China. 
We sincerely thank the anonymous referee whose very constructive comments largely improved the quality of the paper.
This work was supported by 
the Bureau of International Cooperation, Chinese Academy of Sciences under the grant GJHZ1864. 
SZ acknowledges the National Key R\&D Program of China No. 2019YFA0405501. 
AD and HY also acknowledge partial support from the INFN grant InDark and from the Italian Ministry of Education, University and Research (MIUR) under the {\it Departments of Excellence} grant L.232/2016.
This work has made use of data from the European Space Agency (ESA) mission
{\it Gaia} (\url{https://www.cosmos.esa.int/gaia}), processed by the {\it Gaia}
Data Processing and Analysis Consortium (DPAC,
\url{https://www.cosmos.esa.int/web/gaia/dpac/consortium}). Funding for the DPAC
has been provided by national institutions, in particular the institutions
participating in the {\it Gaia} Multilateral Agreement. This research has made use of NASA's Astrophysics Data System Bibliographic Services.

\bibliography{sc}

\begin{thebibliography}{}
\expandafter\ifx\csname natexlab\endcsname\relax\def\natexlab#1{#1}\fi
\providecommand{\url}[1]{\href{#1}{#1}}
\providecommand{\dodoi}[1]{doi:~\href{http://doi.org/#1}{\nolinkurl{#1}}}
\providecommand{\doeprint}[1]{\href{http://ascl.net/#1}{\nolinkurl{http://ascl.net/#1}}}
\providecommand{\doarXiv}[1]{\href{https://arxiv.org/abs/#1}{\nolinkurl{https://arxiv.org/abs/#1}}}

\bibitem[{{Astropy Collaboration} {et~al.}(2013){Astropy Collaboration},
  {Robitaille}, {Tollerud}, {Greenfield}, {Droettboom}, {Bray}, {Aldcroft},
  {Davis}, {Ginsburg}, {Price-Whelan}, {Kerzendorf}, {Conley}, {Crighton},
  {Barbary}, {Muna}, {Ferguson}, {Grollier}, {Parikh}, {Nair}, {Unther},
  {Deil}, {Woillez}, {Conseil}, {Kramer}, {Turner}, {Singer}, {Fox}, {Weaver},
  {Zabalza}, {Edwards}, {Azalee Bostroem}, {Burke}, {Casey}, {Crawford},
  {Dencheva}, {Ely}, {Jenness}, {Labrie}, {Lim}, {Pierfederici}, {Pontzen},
  {Ptak}, {Refsdal}, {Servillat}, \& {Streicher}}]{2013A&A...558A..33A}
{Astropy Collaboration}, {Robitaille}, T.~P., {Tollerud}, E.~J., {et~al.} 2013,
  \aap, 558, A33, \dodoi{10.1051/0004-6361/201322068}

\bibitem[{{Balaguer-N{\'u}{\~n}ez} {et~al.}(2004){Balaguer-N{\'u}{\~n}ez},
  {Jordi}, {Galad{\'{\i}}-Enr{\'{\i}}quez}, \& {Zhao}}]{2004Balaguer}
{Balaguer-N{\'u}{\~n}ez}, L., {Jordi}, C., {Galad{\'{\i}}-Enr{\'{\i}}quez}, D.,
  \& {Zhao}, J.~L. 2004, \aap, 426, 819, \dodoi{10.1051/0004-6361:20041332}

\bibitem[{{Bastian} {et~al.}(2010){Bastian}, {Covey}, \& {Meyer}}]{2010Bastian}
{Bastian}, N., {Covey}, K.~R., \& {Meyer}, M.~R. 2010, \araa, 48, 339,
  \dodoi{10.1146/annurev-astro-082708-101642}

\bibitem[{{Bragg} \& {Kenyon}(2005)}]{2005Bragg}
{Bragg}, A.~E., \& {Kenyon}, S.~J. 2005, \aj, 130, 134, \dodoi{10.1086/430455}

\bibitem[{{Cabrera-Cano} \& {Alfaro}(1990)}]{1990Cabrera}
{Cabrera-Cano}, J., \& {Alfaro}, E.~J. 1990, \aap, 235, 94

\bibitem[{{Cantat-Gaudin} {et~al.}(2018){Cantat-Gaudin}, {Jordi}, {Vallenari},
  {Bragaglia}, {Balaguer-N{\'u}{\~n}ez}, {Soubiran}, {Bossini}, {Moitinho},
  {Castro-Ginard}, {Krone-Martins}, {Casamiquela}, {Sordo}, \&
  {Carrera}}]{GAIA2}
{Cantat-Gaudin}, T., {Jordi}, C., {Vallenari}, A., {et~al.} 2018, \aap, 618,
  A93, \dodoi{10.1051/0004-6361/201833476}

\bibitem[{{Castro-Ginard} {et~al.}(2018){Castro-Ginard}, {Jordi}, {Luri},
  {Julbe}, {Morvan}, {Balaguer-N{\'u}{\~n}ez}, \& {Cantat-
  Gaudin}}]{2018Castro}
{Castro-Ginard}, A., {Jordi}, C., {Luri}, X., {et~al.} 2018, \aap, 618, A59,
  \dodoi{10.1051/0004-6361/201833390}

\bibitem[{{Cui} {et~al.}(2012){Cui}, {Zhao}, {Chu}, {Li}, {Li}, {Zhang}, {Su},
  {Yao}, {Wang}, {Xing}, {Li}, {Zhu}, {Wang}, {Gu}, {Luo}, {Xu}, {Zhang},
  {Liu}, {Zhang}, {Yang}, {Cao}, {Chen}, {Chen}, {Chen}, {Chen}, {Chu}, {Feng},
  {Gong}, {Hou}, {Hu}, {Hu}, {Hu}, {Jia}, {Jiang}, {Jiang}, {Jiang}, {Jin},
  {Li}, {Li}, {Li}, {Liu}, {Liu}, {Lu}, {Mao}, {Men}, {Qi}, {Qi}, {Shi},
  {Tang}, {Tao}, {Wang}, {Wang}, {Wang}, {Wang}, {Wang}, {Wang}, {Wang},
  {Wang}, {Wang}, {Wang}, {Wang}, {Wang}, {Xu}, {Xu}, {Yang}, {Yu}, {Yuan},
  {Yuan}, {Zhai}, {Zhang}, {Zhang}, {Zhang}, {Zhao}, {Zhou}, {Zhou}, {Zhu}, \&
  {Zou}}]{2012Lamost}
{Cui}, X.-Q., {Zhao}, Y.-H., {Chu}, Y.-Q., {et~al.} 2012, Research in Astronomy
  and Astrophysics, 12, 1197, \dodoi{10.1088/1674-4527/12/9/003}

\bibitem[{{Currie} {et~al.}(2010){Currie}, {Hernandez}, {Irwin}, {Kenyon},
  {Tokarz}, {Balog}, {Bragg}, {Berlind}, \& {Calkins}}]{2010Currie}
{Currie}, T., {Hernandez}, J., {Irwin}, J., {et~al.} 2010, The Astrophysical
  Journal Supplement Series, 186, 191, \dodoi{10.1088/0067-0049/186/2/191}

\bibitem[{{Deacon} \& {Hambly}(2004)}]{2004Deacon}
{Deacon}, N.~R., \& {Hambly}, N.~C. 2004, \aap, 416, 125,
  \dodoi{10.1051/0004-6361:20034238}

\bibitem[{{Diaferio}(1999)}]{1999Diaferio}
{Diaferio}, A. 1999, \mnras, 309, 610, \dodoi{10.1046/j.1365-8711.1999.02864.x}

\bibitem[{{Dias} {et~al.}(2002){Dias}, {Alessi}, {Moitinho}, \&
  {L{\'e}pine}}]{2002Dias}
{Dias}, W.~S., {Alessi}, B.~S., {Moitinho}, A., \& {L{\'e}pine}, J.~R.~D. 2002,
  \aap, 389, 871, \dodoi{10.1051/0004-6361:20020668}

\bibitem[{{Dias} {et~al.}(2006){Dias}, {Assafin}, {Fl{\'o}rio}, {Alessi}, \&
  {L{\'{\i}}bero}}]{2006Dias}
{Dias}, W.~S., {Assafin}, M., {Fl{\'o}rio}, V., {Alessi}, B.~S., \&
  {L{\'{\i}}bero}, V. 2006, \aap, 446, 949, \dodoi{10.1051/0004-6361:20052741}

\bibitem[{Ester {et~al.}(1996)Ester, Kriegel, Sander, \& Xu}]{DBSCAN}
Ester, M., Kriegel, H.-P., Sander, J., \& Xu, X. 1996, in Proceedings of the
  Second International Conference on Knowledge Discovery and Data Mining,
  KDD'96 (AAAI Press), 226--231.
\newblock \url{http://dl.acm.org/citation.cfm?id=3001460.3001507}

\bibitem[{Everitt {et~al.}(2011)Everitt, Landau, Leese, \& Stahl}]{everitt2011}
Everitt, B.~S., Landau, S., Leese, M., \& Stahl, D. 2011, Cluster Analysis, 5th
  Edition (Wiley Online Library), 71--110

\bibitem[{{Gaia Collaboration} {et~al.}(2016){Gaia Collaboration}, {Prusti},
  {de Bruijne}, {Brown}, {Vallenari}, \& et~al.}]{2016Gaia}
{Gaia Collaboration}, {Prusti}, T., {de Bruijne}, J.~H.~J., {et~al.} 2016,
  \aap, 595, A1, \dodoi{10.1051/0004-6361/201629272}

\bibitem[{{Gaia Collaboration} {et~al.}(2018{\natexlab{a}}){Gaia
  Collaboration}, {Brown}, {Vallenari}, {Prusti}, {de Bruijne}, {Babusiaux},
  {Bailer-Jones}, {Biermann}, {Evans}, {Eyer}, {Jansen}, {Jordi}, {Klioner},
  {Lammers}, {Lindegren}, {Luri}, {Mignard}, {Panem}, {Pourbaix}, {Randich},
  {Sartoretti}, {Siddiqui}, {Soubiran}, {van Leeuwen}, {Walton}, {Arenou},
  {Bastian}, {Cropper}, {Drimmel}, {Katz}, {Lattanzi}, {Bakker}, {Cacciari},
  {Casta{\~n}eda}, {Chaoul}, {Cheek}, {De Angeli}, {Fabricius}, {Guerra},
  {Holl}, {Masana}, {Messineo}, {Mowlavi}, {Nienartowicz}, {Panuzzo},
  {Portell}, {Riello}, {Seabroke}, {Tanga}, {Th{\'e}venin}, {Gracia-Abril},
  {Comoretto}, {Garcia-Reinaldos}, {Teyssier}, {Altmann}, {Andrae}, {Audard},
  {Bellas-Velidis}, {Benson}, {Berthier}, {Blomme}, {Burgess}, {Busso},
  {Carry}, {Cellino}, {Clementini}, {Clotet}, {Creevey}, {Davidson}, {De
  Ridder}, {Delchambre}, {Dell'Oro}, {Ducourant},
  {Fern{\'a}ndez-Hern{\'a}ndez}, {Fouesneau}, {Fr{\'e}mat}, {Galluccio},
  {Garc{\'\i}a-Torres}, {Gonz{\'a}lez-N{\'u}{\~n}ez}, {Gonz{\'a}lez-Vidal},
  {Gosset}, {Guy}, {Halbwachs}, {Hambly}, {Harrison}, {Hern{\'a}ndez},
  {Hestroffer}, {Hodgkin}, {Hutton}, {Jasniewicz}, {Jean-Antoine-Piccolo},
  {Jordan}, {Korn}, {Krone-Martins}, {Lanzafame}, {Lebzelter}, {L{\"o}ffler},
  {Manteiga}, {Marrese}, {Mart{\'\i}n-Fleitas}, {Moitinho}, {Mora}, {Muinonen},
  {Osinde}, {Pancino}, {Pauwels}, {Petit}, {Recio-Blanco}, {Richards},
  {Rimoldini}, {Robin}, {Sarro}, {Siopis}, {Smith}, {Sozzetti}, {S{\"u}veges},
  {Torra}, {van Reeven}, {Abbas}, {Abreu Aramburu}, {Accart}, {Aerts},
  {Altavilla}, {{\'A}lvarez}, {Alvarez}, {Alves}, {Anderson}, {Andrei},
  {Anglada Varela}, {Antiche}, {Antoja}, {Arcay}, {Astraatmadja}, {Bach},
  {Baker}, {Balaguer-N{\'u}{\~n}ez}, {Balm}, {Barache}, {Barata}, {Barbato},
  {Barblan}, {Barklem}, {Barrado}, {Barros}, {Barstow}, {Bartholom{\'e}
  Mu{\~n}oz}, {Bassilana}, {Becciani}, {Bellazzini}, {Berihuete}, {Bertone},
  {Bianchi}, {Bienaym{\'e}}, {Blanco-Cuaresma}, {Boch}, {Boeche}, {Bombrun},
  {Borrachero}, {Bossini}, {Bouquillon}, {Bourda}, {Bragaglia}, {Bramante},
  {Breddels}, {Bressan}, {Brouillet}, {Br{\"u}semeister}, {Brugaletta},
  {Bucciarelli}, {Burlacu}, {Busonero}, {Butkevich}, {Buzzi}, {Caffau},
  {Cancelliere}, {Cannizzaro}, {Cantat-Gaudin}, {Carballo}, {Carlucci},
  {Carrasco}, {Casamiquela}, {Castellani}, {Castro-Ginard}, {Charlot},
  {Chemin}, {Chiavassa}, {Cocozza}, {Costigan}, {Cowell}, {Crifo}, {Crosta},
  {Crowley}, {Cuypers}, {Dafonte}, {Damerdji}, {Dapergolas}, {David}, {David},
  {de Laverny}, {De Luise}, {De March}, {de Martino}, {de Souza}, {de Torres},
  {Debosscher}, {del Pozo}, {Delbo}, {Delgado}, {Delgado}, {Di Matteo},
  {Diakite}, {Diener}, {Distefano}, {Dolding}, {Drazinos}, {Dur{\'a}n},
  {Edvardsson}, {Enke}, {Eriksson}, {Esquej}, {Eynard Bontemps}, {Fabre},
  {Fabrizio}, {Faigler}, {Falc{\~a}o}, {Farr{\`a}s Casas}, {Federici},
  {Fedorets}, {Fernique}, {Figueras}, {Filippi}, {Findeisen}, {Fonti},
  {Fraile}, {Fraser}, {Fr{\'e}zouls}, {Gai}, {Galleti}, {Garabato},
  {Garc{\'\i}a-Sedano}, {Garofalo}, {Garralda}, {Gavel}, {Gavras}, {Gerssen},
  {Geyer}, {Giacobbe}, {Gilmore}, {Girona}, {Giuffrida}, {Glass}, {Gomes},
  {Granvik}, {Gueguen}, {Guerrier}, {Guiraud}, {Guti{\'e}rrez-S{\'a}nchez},
  {Haigron}, {Hatzidimitriou}, {Hauser}, {Haywood}, {Heiter}, {Helmi}, {Heu},
  {Hilger}, {Hobbs}, {Hofmann}, {Holland}, {Huckle}, {Hypki}, {Icardi},
  {Jan{\ss}en}, {Jevardat de Fombelle}, {Jonker}, {Juh{\'a}sz}, {Julbe},
  {Karampelas}, {Kewley}, {Klar}, {Kochoska}, {Kohley}, {Kolenberg},
  {Kontizas}, {Kontizas}, {Koposov}, {Kordopatis}, {Kostrzewa-Rutkowska},
  {Koubsky}, {Lambert}, {Lanza}, {Lasne}, {Lavigne}, {Le Fustec}, {Le
  Poncin-Lafitte}, {Lebreton}, {Leccia}, {Leclerc}, {Lecoeur-Taibi},
  {Lenhardt}, {Leroux}, {Liao}, {Licata}, {Lindstr{\o}m}, {Lister}, {Livanou},
  {Lobel}, {L{\'o}pez}, {Managau}, {Mann}, {Mantelet}, {Marchal}, {Marchant},
  {Marconi}, {Marinoni}, {Marschalk{\'o}}, {Marshall}, {Martino}, {Marton},
  {Mary}, {Massari}, {Matijevi{\v{c}}}, {Mazeh}, {McMillan}, {Messina},
  {Michalik}, {Millar}, {Molina}, {Molinaro}, {Moln{\'a}r}, {Montegriffo},
  {Mor}, {Morbidelli}, {Morel}, {Morris}, {Mulone}, {Muraveva}, {Musella},
  {Nelemans}, {Nicastro}, {Noval}, {O'Mullane}, {Ord{\'e}novic},
  {Ord{\'o}{\~n}ez-Blanco}, {Osborne}, {Pagani}, {Pagano}, {Pailler},
  {Palacin}, {Palaversa}, {Panahi}, {Pawlak}, {Piersimoni}, {Pineau}, {Plachy},
  {Plum}, {Poggio}, {Poujoulet}, {Pr{\v{s}}a}, {Pulone}, {Racero}, {Ragaini},
  {Rambaux}, {Ramos-Lerate}, {Regibo}, {Reyl{\'e}}, {Riclet}, {Ripepi}, {Riva},
  {Rivard}, {Rixon}, {Roegiers}, {Roelens}, {Romero-G{\'o}mez}, {Rowell},
  {Royer}, {Ruiz-Dern}, {Sadowski}, {Sagrist{\`a} Sell{\'e}s}, {Sahlmann},
  {Salgado}, {Salguero}, {Sanna}, {Santana-Ros}, {Sarasso}, {Savietto},
  {Schultheis}, {Sciacca}, {Segol}, {Segovia}, {S{\'e}gransan}, {Shih},
  {Siltala}, {Silva}, {Smart}, {Smith}, {Solano}, {Solitro}, {Sordo}, {Soria
  Nieto}, {Souchay}, {Spagna}, {Spoto}, {Stampa}, {Steele},
  {Steidelm{\"u}ller}, {Stephenson}, {Stoev}, {Suess}, {Surdej}, {Szabados},
  {Szegedi-Elek}, {Tapiador}, {Taris}, {Tauran}, {Taylor}, {Teixeira},
  {Terrett}, {Teyssand ier}, {Thuillot}, {Titarenko}, {Torra Clotet}, {Turon},
  {Ulla}, {Utrilla}, {Uzzi}, {Vaillant}, {Valentini}, {Valette}, {van Elteren},
  {Van Hemelryck}, {van Leeuwen}, {Vaschetto}, {Vecchiato}, {Veljanoski},
  {Viala}, {Vicente}, {Vogt}, {von Essen}, {Voss}, {Votruba}, {Voutsinas},
  {Walmsley}, {Weiler}, {Wertz}, {Wevers}, {Wyrzykowski}, {Yoldas},
  {{\v{Z}}erjal}, {Ziaeepour}, {Zorec}, {Zschocke}, {Zucker}, {Zurbach}, \&
  {Zwitter}}]{2018Gaia}
{Gaia Collaboration}, {Brown}, A.~G.~A., {Vallenari}, A., {et~al.}
  2018{\natexlab{a}}, Astronomy and Astrophysics, 616, A1,
  \dodoi{10.1051/0004-6361/201833051}

\bibitem[{{Gaia Collaboration} {et~al.}(2018{\natexlab{b}}){Gaia
  Collaboration}, {Brown}, {Vallenari}, {Prusti}, {de Bruijne}, {Babusiaux},
  {Bailer-Jones}, {Biermann}, {Evans}, {Eyer}, {Jansen}, {Jordi}, \&
  {Klioner}}]{2018Gaiadr2}
---. 2018{\natexlab{b}}, \aap, 616, A1, \dodoi{10.1051/0004-6361/201833051}

\bibitem[{{Gaia Collaboration} {et~al.}(2018{\natexlab{c}}){Gaia
  Collaboration}, {Babusiaux}, {van Leeuwen}, {Barstow}, {Jordi}, {Vallenari},
  {Bossini}, {Bressan}, {Cantat-Gaudin}, {van Leeuwen}, {Brown}, {Prusti}, {de
  Bruijne}, {Bailer-Jones}, {Biermann}, {Evans}, {Eyer}, {Jansen}, {Klioner},
  {Lammers}, {Lindegren}, {Luri}, {Mignard}, {Panem}, {Pourbaix}, {Randich},
  {Sartoretti}, {Siddiqui}, {Soubiran}, {Walton}, {Arenou}, {Bastian},
  {Cropper}, {Drimmel}, {Katz}, {Lattanzi}, {Bakker}, {Cacciari},
  {Casta{\~n}eda}, {Chaoul}, {Cheek}, {De Angeli}, {Fabricius}, {Guerra},
  {Holl}, {Masana}, {Messineo}, {Mowlavi}, {Nienartowicz}, {Panuzzo},
  {Portell}, {Riello}, {Seabroke}, {Tanga}, {Th{\'e}venin}, {Gracia-Abril},
  {Comoretto}, {Garcia-Reinaldos}, {Teyssier}, {Altmann}, {Andrae}, {Audard},
  {Bellas-Velidis}, {Benson}, {Berthier}, {Blomme}, {Burgess}, {Busso},
  {Carry}, {Cellino}, {Clementini}, {Clotet}, {Creevey}, {Davidson}, {De
  Ridder}, {Delchambre}, {Dell'Oro}, {Ducourant},
  {Fern{\'a}ndez-Hern{\'a}ndez}, {Fouesneau}, {Fr{\'e}mat}, {Galluccio},
  {Garc{\'\i}a-Torres}, {Gonz{\'a}lez-N{\'u}{\~n}ez}, {Gonz{\'a}lez-Vidal},
  {Gosset}, {Guy}, {Halbwachs}, {Hambly}, {Harrison}, {Hern{\'a}ndez},
  {Hestroffer}, {Hodgkin}, {Hutton}, {Jasniewicz}, {Jean-Antoine-Piccolo},
  {Jordan}, {Korn}, {Krone-Martins}, {Lanzafame}, {Lebzelter}, {L{\"o}ffler},
  {Manteiga}, {Marrese}, {Mart{\'\i}n-Fleitas}, {Moitinho}, {Mora}, {Muinonen},
  {Osinde}, {Pancino}, {Pauwels}, {Petit}, {Recio-Blanco}, {Richards},
  {Rimoldini}, {Robin}, {Sarro}, {Siopis}, {Smith}, {Sozzetti}, {S{\"u}veges},
  {Torra}, {van Reeven}, {Abbas}, {Abreu Aramburu}, {Accart}, {Aerts},
  {Altavilla}, {{\'A}lvarez}, {Alvarez}, {Alves}, {Anderson}, {Andrei},
  {Anglada Varela}, {Antiche}, {Antoja}, {Arcay}, {Astraatmadja}, {Bach},
  {Baker}, {Balaguer-N{\'u}{\~n}ez}, {Balm}, {Barache}, {Barata}, {Barbato},
  {Barblan}, {Barklem}, {Barrado}, {Barros}, {Bartholom{\'e} Mu{\~n}oz},
  {Bassilana}, {Becciani}, {Bellazzini}, {Berihuete}, {Bertone}, {Bianchi},
  {Bienaym{\'e}}, {Blanco-Cuaresma}, {Boch}, {Boeche}, {Bombrun}, {Borrachero},
  {Bouquillon}, {Bourda}, {Bragaglia}, {Bramante}, {Breddels}, {Brouillet},
  {Br{\"u}semeister}, {Brugaletta}, {Bucciarelli}, {Burlacu}, {Busonero},
  {Butkevich}, {Buzzi}, {Caffau}, {Cancelliere}, {Cannizzaro}, {Carballo},
  {Carlucci}, {Carrasco}, {Casamiquela}, {Castellani}, {Castro-Ginard},
  {Charlot}, {Chemin}, {Chiavassa}, {Cocozza}, {Costigan}, {Cowell}, {Crifo},
  {Crosta}, {Crowley}, {Cuypers}, {Dafonte}, {Damerdji}, {Dapergolas}, {David},
  {David}, {de Laverny}, {De Luise}, {De March}, {de Martino}, {de Souza}, {de
  Torres}, {Debosscher}, {del Pozo}, {Delbo}, {Delgado}, {Delgado}, {Diakite},
  {Diener}, {Distefano}, {Dolding}, {Drazinos}, {Dur{\'a}n}, {Edvardsson},
  {Enke}, {Eriksson}, {Esquej}, {Eynard Bontemps}, {Fabre}, {Fabrizio},
  {Faigler}, {Falc{\~a}o}, {Farr{\`a}s Casas}, {Federici}, {Fedorets},
  {Fernique}, {Figueras}, {Filippi}, {Findeisen}, {Fonti}, {Fraile}, {Fraser},
  {Fr{\'e}zouls}, {Gai}, {Galleti}, {Garabato}, {Garc{\'\i}a-Sedano},
  {Garofalo}, {Garralda}, {Gavel}, {Gavras}, {Gerssen}, {Geyer}, {Giacobbe},
  {Gilmore}, {Girona}, {Giuffrida}, {Glass}, {Gomes}, {Granvik}, {Gueguen},
  {Guerrier}, {Guiraud}, {Guti{\'e}}, {Haigron}, {Hatzidimitriou}, {Hauser},
  {Haywood}, {Heiter}, {Helmi}, {Heu}, {Hilger}, {Hobbs}, {Hofmann}, {Holland},
  {Huckle}, {Hypki}, {Icardi}, {Jan{\ss}en}, {Jevardat de Fombelle}, {Jonker},
  {Juh{\'a}sz}, {Julbe}, {Karampelas}, {Kewley}, {Klar}, {Kochoska}, {Kohley},
  {Kolenberg}, {Kontizas}, {Kontizas}, {Koposov}, {Kordopatis},
  {Kostrzewa-Rutkowska}, {Koubsky}, {Lambert}, {Lanza}, {Lasne}, {Lavigne}, {Le
  Fustec}, {Le Poncin-Lafitte}, {Lebreton}, {Leccia}, {Leclerc},
  {Lecoeur-Taibi}, {Lenhardt}, {Leroux}, {Liao}, {Licata}, {Lindstr{\o}m},
  {Lister}, {Livanou}, {Lobel}, {L{\'o}pez}, {Managau}, {Mann}, {Mantelet},
  {Marchal}, {Marchant}, {Marconi}, {Marinoni}, {Marschalk{\'o}}, {Marshall},
  {Martino}, {Marton}, {Mary}, {Massari}, {Matijevi{\v{c}}}, {Mazeh},
  {McMillan}, {Messina}, {Michalik}, {Millar}, {Molina}, {Molinaro},
  {Moln{\'a}r}, {Montegriffo}, {Mor}, {Morbidelli}, {Morel}, {Morris},
  {Mulone}, {Muraveva}, {Musella}, {Nelemans}, {Nicastro}, {Noval},
  {O'Mullane}, {Ord{\'e}novic}, {Ord{\'o}{\~n}ez-Blanco}, {Osborne}, {Pagani},
  {Pagano}, {Pailler}, {Palacin}, {Palaversa}, {Panahi}, {Pawlak},
  {Piersimoni}, {Pineau}, {Plachy}, {Plum}, {Poggio}, {Poujoulet},
  {Pr{\v{s}}a}, {Pulone}, {Racero}, {Ragaini}, {Rambaux}, {Ramos-Lerate},
  {Regibo}, {Reyl{\'e}}, {Riclet}, {Ripepi}, {Riva}, {Rivard}, {Rixon},
  {Roegiers}, {Roelens}, {Romero-G{\'o}mez}, {Rowell}, {Royer}, {Ruiz-Dern},
  {Sadowski}, {Sagrist{\`a} Sell{\'e}s}, {Sahlmann}, {Salgado}, {Salguero},
  {Sanna}, {Santana-Ros}, {Sarasso}, {Savietto}, {Schultheis}, {Sciacca},
  {Segol}, {Segovia}, {S{\'e}gransan}, {Shih}, {Siltala}, {Silva}, {Smart},
  {Smith}, {Solano}, {Solitro}, {Sordo}, {Soria Nieto}, {Souchay}, {Spagna},
  {Spoto}, {Stampa}, {Steele}, {Steidelm{\"u}ller}, {Stephenson}, {Stoev},
  {Suess}, {Surdej}, {Szabados}, {Szegedi-Elek}, {Tapiador}, {Taris}, {Tauran},
  {Taylor}, {Teixeira}, {Terrett}, {Teyssand ier}, {Thuillot}, {Titarenko},
  {Torra Clotet}, {Turon}, {Ulla}, {Utrilla}, {Uzzi}, {Vaillant}, {Valentini},
  {Valette}, {van Elteren}, {Van Hemelryck}, {Vaschetto}, {Vecchiato},
  {Veljanoski}, {Viala}, {Vicente}, {Vogt}, {von Essen}, {Voss}, {Votruba},
  {Voutsinas}, {Walmsley}, {Weiler}, {Wertz}, {Wevers}, {Wyrzykowski},
  {Yoldas}, {{\v{Z}}erjal}, {Ziaeepour}, {Zorec}, {Zschocke}, {Zucker},
  {Zurbach}, \& {Zwitter}}]{gaia2018b}
{Gaia Collaboration}, {Babusiaux}, C., {van Leeuwen}, F., {et~al.}
  2018{\natexlab{c}}, \aap, 616, A10, \dodoi{10.1051/0004-6361/201832843}

\bibitem[{{Gao}(2014)}]{2014Gao}
{Gao}, X.-H. 2014, Research in Astronomy and Astrophysics, 14, 159,
  \dodoi{10.1088/1674-4527/14/2/004}

\bibitem[{Ivezi{\'c} {et~al.}(2014)Ivezi{\'c}, Connolly, VanderPlas, \&
  Gray}]{ivezic2014statistics}
Ivezi{\'c}, {\v{Z}}., Connolly, A., VanderPlas, J., \& Gray, A. 2014,
  Statistics, Data Mining, and Machine Learning in Astronomy: A Practical
  Python Guide for the Analysis of Survey Data, Princeton Series in Modern
  Observational Astronomy (Princeton University Press).
\newblock \url{https://books.google.com/books?id=2fM8AQAAQBAJ}

\bibitem[{{Javakhishvili} {et~al.}(2006){Javakhishvili}, {Kukhianidze},
  {Todua}, \& {Inasaridze}}]{2006Javakhishvili}
{Javakhishvili}, G., {Kukhianidze}, V., {Todua}, M., \& {Inasaridze}, R. 2006,
  \aap, 447, 915, \dodoi{10.1051/0004-6361:20040297}

\bibitem[{{Keller} {et~al.}(2001){Keller}, {Grebel}, {Miller}, \&
  {Yoss}}]{2001Keller}
{Keller}, S.~C., {Grebel}, E.~K., {Miller}, G.~J., \& {Yoss}, K.~M. 2001, \aj,
  122, 248, \dodoi{10.1086/321139}

\bibitem[{{Kharchenko} {et~al.}(2004){Kharchenko}, {Piskunov}, {R{\"o}ser},
  {Schilbach}, \& {Scholz}}]{2004Kharchenko}
{Kharchenko}, N.~V., {Piskunov}, A.~E., {R{\"o}ser}, S., {Schilbach}, E., \&
  {Scholz}, R.-D. 2004, Astronomische Nachrichten, 325, 740,
  \dodoi{10.1002/asna.200410256}

\bibitem[{{Kharchenko} {et~al.}(2013){Kharchenko}, {Piskunov}, {Schilbach},
  {R{\"o}ser}, \& {Scholz}}]{2013Kharchenko}
{Kharchenko}, N.~V., {Piskunov}, A.~E., {Schilbach}, E., {R{\"o}ser}, S., \&
  {Scholz}, R.~D. 2013, \aap, 558, A53, \dodoi{10.1051/0004-6361/201322302}

\bibitem[{{King}(1962)}]{1962King}
{King}, I. 1962, \aj, 67, 471, \dodoi{10.1086/108756}

\bibitem[{{Kordopatis} {et~al.}(2013){Kordopatis}, {Gilmore}, {Steinmetz},
  {Boeche}, {Seabroke}, {Siebert}, {Zwitter}, {Binney}, {de Laverny},
  {Recio-Blanco}, {Williams}, {Piffl}, {Enke}, {Roeser}, {Bijaoui}, {Wyse},
  {Freeman}, {Munari}, {Carrillo}, {Anguiano}, {Burton}, {Campbell}, {Cass},
  {Fiegert}, {Hartley}, {Parker}, {Reid}, {Ritter}, {Russell}, {Stupar},
  {Watson}, {Bienaym{\'e}}, {Bland -Hawthorn}, {Gerhard}, {Gibson}, {Grebel},
  {Helmi}, {Navarro}, {Conrad}, {Famaey}, {Faure}, {Just}, {Kos},
  {Matijevi{\v{c}}}, {McMillan}, {Minchev}, {Scholz}, {Sharma}, {Siviero}, {de
  Boer}, \& {{\v{Z}}erjal}}]{2013Kordopatis}
{Kordopatis}, G., {Gilmore}, G., {Steinmetz}, M., {et~al.} 2013, \aj, 146, 134,
  \dodoi{10.1088/0004-6256/146/5/134}

\bibitem[{{Kozhurina-Platais} {et~al.}(1995){Kozhurina-Platais}, {Girard},
  {Platais}, {van Altena}, {Ianna}, \& {Cannon}}]{1995Kozhurina-Platais}
{Kozhurina-Platais}, V., {Girard}, T.~M., {Platais}, I., {et~al.} 1995, \aj,
  109, 672, \dodoi{10.1086/117310}

\bibitem[{{Krone-Martins} \& {Moitinho}(2014)}]{2014Krone}
{Krone-Martins}, A., \& {Moitinho}, A. 2014, \aap, 561, A57,
  \dodoi{10.1051/0004-6361/201321143}

\bibitem[{{Krone-Martins} {et~al.}(2010){Krone-Martins}, {Soubiran},
  {Ducourant}, {Teixeira}, \& {Le Campion}}]{2010Krone-Martins}
{Krone-Martins}, A., {Soubiran}, C., {Ducourant}, C., {Teixeira}, R., \& {Le
  Campion}, J.~F. 2010, \aap, 516, A3, \dodoi{10.1051/0004-6361/200913881}

\bibitem[{{Li} {et~al.}(2019){Li}, {Sun}, {de Grijs}, {Deng}, {Wang},
  {Cordoni}, \& {Milone}}]{2019Li}
{Li}, C., {Sun}, W., {de Grijs}, R., {et~al.} 2019, The Astrophysical Journal,
  876, 65, \dodoi{10.3847/1538-4357/ab15d2}

\bibitem[{Liu {et~al.}(2018)Liu, Yu, Diaferio, Tozzi, Hwang, Umetsu, Okabe, \&
  Yang}]{Liu2018b}
Liu, A., Yu, H., Diaferio, A., {et~al.} 2018, The Astrophysical Journal, 863

\bibitem[{{Majewski} {et~al.}(2017){Majewski}, {Schiavon}, {Frinchaboy},
  {Allende Prieto}, {Barkhouser}, {Bizyaev}, {Blank}, {Brunner}, {Burton},
  {Carrera}, {Chojnowski}, {Cunha}, {Epstein}, {Fitzgerald}, {Garc{\'\i}a
  P{\'e}rez}, {Hearty}, {Henderson}, {Holtzman}, {Johnson}, {Lam}, {Lawler},
  {Maseman}, {M{\'e}sz{\'a}ros}, {Nelson}, {Nguyen}, {Nidever}, {Pinsonneault},
  {Shetrone}, {Smee}, {Smith}, {Stolberg}, {Skrutskie}, {Walker}, {Wilson},
  {Zasowski}, {Anders}, {Basu}, {Beland}, {Blanton}, {Bovy}, {Brownstein},
  {Carlberg}, {Chaplin}, {Chiappini}, {Eisenstein}, {Elsworth}, {Feuillet},
  {Fleming}, {Galbraith-Frew}, {Garc{\'\i}a}, {Garc{\'\i}a-Hern{\'a}ndez},
  {Gillespie}, {Girardi}, {Gunn}, {Hasselquist}, {Hayden}, {Hekker}, {Ivans},
  {Kinemuchi}, {Klaene}, {Mahadevan}, {Mathur}, {Mosser}, {Muna}, {Munn},
  {Nichol}, {O'Connell}, {Parejko}, {Robin}, {Rocha-Pinto}, {Schultheis},
  {Serenelli}, {Shane}, {Silva Aguirre}, {Sobeck}, {Thompson}, {Troup},
  {Weinberg}, \& {Zamora}}]{2017Majewski}
{Majewski}, S.~R., {Schiavon}, R.~P., {Frinchaboy}, P.~M., {et~al.} 2017, \aj,
  154, 94, \dodoi{10.3847/1538-3881/aa784d}

\bibitem[{{Malbet} {et~al.}(2019){Malbet}, {Abbas}, {Alves}, {Boehm}, {Brown},
  {Chemin}, {Correia}, {Courbin}, {Darling}, {Diaferio}, {Fortin}, {Fridlund},
  {Gnedin}, {Holl}, {Krone-Martins}, {L{\'e}ger}, {Labadie}, {Laskar}, {Mamon},
  {McArthur}, {Michalik}, {Moitinho}, {Oertel}, {Ostorero}, {Schneider},
  {Scott}, {Shao}, {Sozzetti}, {Tomsick}, {Valluri}, \& {Wyse}}]{2019Malbet}
{Malbet}, F., {Abbas}, U., {Alves}, J., {et~al.} 2019, arXiv e-prints,
  arXiv:1910.08028.
\newblock \doarXiv{1910.08028}

\bibitem[{{Materne}(1978)}]{1978Materne}
{Materne}, J. 1978, \aap, 63, 401

\bibitem[{{Nambiar} {et~al.}(2019){Nambiar}, {Das}, {Vig}, \&
  {Gorthi}}]{2019Nambiar}
{Nambiar}, S., {Das}, S., {Vig}, S., \& {Gorthi}, R. S.~S. 2019, \mnras, 482,
  3789, \dodoi{10.1093/mnras/sty2851}

\bibitem[{Pedregosa {et~al.}(2011)Pedregosa, Varoquaux, Gramfort, Michel,
  Thirion, Grisel, Blondel, Prettenhofer, Weiss, Dubourg, Vanderplas, Passos,
  Cournapeau, Brucher, Perrot, \& Duchesnay}]{scikit-learn}
Pedregosa, F., Varoquaux, G., Gramfort, A., {et~al.} 2011, Journal of Machine
  Learning Research, 12, 2825

\bibitem[{Ravindran {et~al.}(2006)Ravindran, Ragsdell, \&
  Reklaitis}]{ravindran2006}
Ravindran, A., Ragsdell, K., \& Reklaitis, G. 2006, Engineering Optimization:
  Methods and Applications (Wiley).
\newblock \url{https://books.google.com/books?id=Hf4eAQAAIAAJ}

\bibitem[{{Sampedro} \& {Alfaro}(2016)}]{2016Sampedro}
{Sampedro}, L., \& {Alfaro}, E.~J. 2016, \mnras, 457, 3949,
  \dodoi{10.1093/mnras/stw243}

\bibitem[{{Sanders}(1971)}]{1971Sanders}
{Sanders}, W.~L. 1971, \aap, 14, 226

\bibitem[{{Sarro} {et~al.}(2014){Sarro}, {Bouy}, {Berihuete}, {Bertin},
  {Moraux}, {Bouvier}, {Cuillandre}, {Barrado}, \& {Solano}}]{2014Sarro}
{Sarro}, L.~M., {Bouy}, H., {Berihuete}, A., {et~al.} 2014, \aap, 563, A45,
  \dodoi{10.1051/0004-6361/201322413}

\bibitem[{{Schmeja}(2011)}]{2011Schmeja}
{Schmeja}, S. 2011, Astronomische Nachrichten, 332, 172,
  \dodoi{10.1002/asna.201011484}

\bibitem[{Serna \& Gerbal(1996)}]{Serna1996}
Serna, A., \& Gerbal, D. 1996, Astronomy and Astrophysics, 309, 65

\bibitem[{{Serra} \& {Diaferio}(2013)}]{2013Serra}
{Serra}, A.~L., \& {Diaferio}, A. 2013, \apj, 768, 116,
  \dodoi{10.1088/0004-637X/768/2/116}

\bibitem[{{Serra} {et~al.}(2011){Serra}, {Diaferio}, {Murante}, \&
  {Borgani}}]{Serra2011}
{Serra}, A.~L., {Diaferio}, A., {Murante}, G., \& {Borgani}, S. 2011, \mnras,
  412, 800, \dodoi{10.1111/j.1365-2966.2010.17946.x}

\bibitem[{{Slesnick} {et~al.}(2002){Slesnick}, {Hillenbrand}, \&
  {Massey}}]{2002Slesnick}
{Slesnick}, C.~L., {Hillenbrand}, L.~A., \& {Massey}, P. 2002, \apj, 576, 880,
  \dodoi{10.1086/341865}

\bibitem[{{The Theia Collaboration} {et~al.}(2017){The Theia Collaboration},
  {Boehm}, {Krone-Martins}, {Amorim}, {Anglada-Escude}, {Brandeker}, {Courbin},
  {Ensslin}, {Falcao}, {Freese}, {Holl}, {Labadie}, {Leger}, {Malbet}, {Mamon},
  {McArthur}, {Mora}, {Shao}, {Sozzetti}, {Spolyar}, {Villaver}, {Albertus},
  {Bertone}, {Bouy}, {Boylan-Kolchin}, {Brown}, {Brown}, {Cardoso}, {Chemin},
  {Claudi}, {Correia}, {Crosta}, {Crouzier}, {Cyr-Racine}, {Damasso}, {da
  Silva}, {Davies}, {Das}, {Dayal}, {de Val-Borro}, {Diaferio}, {Erickcek},
  {Fairbairn}, {Fortin}, {Fridlund}, {Garcia}, {Gnedin}, {Goobar}, {Gordo},
  {Goullioud}, {Hambly}, {Hara}, {Hobbs}, {Hog}, {Holland}, {Ibata}, {Jordi},
  {Klioner}, {Kopeikin}, {Lacroix}, {Laskar}, {Le Poncin-Lafitte}, {Luri},
  {Majumdar}, {Makarov}, {Massey}, {Mennesson}, {Michalik}, {Moitinho de
  Almeida}, {Mourao}, {Moustakas}, {Murray}, {Muterspaugh}, {Oertel},
  {Ostorero}, {Perez-Garcia}, {Platais}, {de Mora}, {Quirrenbach}, {Randall},
  {Read}, {Regos}, {Rory}, {Rybicki}, {Scott}, {Schneider}, {Scholtz},
  {Siebert}, {Tereno}, {Tomsick}, {Traub}, {Valluri}, {Walker}, {Walton},
  {Watkins}, {White}, {Evans}, {Wyrzykowski}, \& {Wyse}}]{Theia2017}
{The Theia Collaboration}, {Boehm}, C., {Krone-Martins}, A., {et~al.} 2017,
  arXiv e-prints, arXiv:1707.01348.
\newblock \doarXiv{1707.01348}

\bibitem[{{Uribe} {et~al.}(2002){Uribe}, {Garc{\'\i}a-Varela},
  {Sabogal-Mart{\'\i}nez}, {Higuera G.}, \& {Brieva}}]{2002Uribe}
{Uribe}, A., {Garc{\'\i}a-Varela}, J.-A., {Sabogal-Mart{\'\i}nez}, B.-E.,
  {Higuera G.}, M.~A., \& {Brieva}, E. 2002, \pasp, 114, 233,
  \dodoi{10.1086/338428}

\bibitem[{{Vasilevskis} {et~al.}(1958){Vasilevskis}, {Klemola}, \&
  {Preston}}]{1958Vasilevskis}
{Vasilevskis}, S., {Klemola}, A., \& {Preston}, G. 1958, \aj, 63, 387,
  \dodoi{10.1086/107787}

\bibitem[{{Yanny} {et~al.}(2009){Yanny}, {Rockosi}, {Newberg}, {Knapp},
  {Adelman-McCarthy}, {Alcorn}, {Allam}, {Allende Prieto}, {An}, {Anderson},
  {Anderson}, {Bailer-Jones}, {Bastian}, {Beers}, {Bell}, {Belokurov},
  {Bizyaev}, {Blythe}, {Bochanski}, {Boroski}, {Brinchmann}, {Brinkmann},
  {Brewington}, {Carey}, {Cudworth}, {Evans}, {Evans}, {Gates}, {G{\"a}nsicke},
  {Gillespie}, {Gilmore}, {Nebot Gomez-Moran}, {Grebel}, {Greenwell}, {Gunn},
  {Jordan}, {Jordan}, {Harding}, {Harris}, {Hendry}, {Holder}, {Ivans},
  {Ivezi{\v{c}}}, {Jester}, {Johnson}, {Kent}, {Kleinman}, {Kniazev},
  {Krzesinski}, {Kron}, {Kuropatkin}, {Lebedeva}, {Lee}, {French Leger},
  {L{\'e}pine}, {Levine}, {Lin}, {Long}, {Loomis}, {Lupton}, {Malanushenko},
  {Malanushenko}, {Margon}, {Martinez-Delgado}, {McGehee}, {Monet}, {Morrison},
  {Munn}, {Neilsen}, {Nitta}, {Norris}, {Oravetz}, {Owen}, {Padmanabhan},
  {Pan}, {Peterson}, {Pier}, {Platson}, {Re Fiorentin}, {Richards}, {Rix},
  {Schlegel}, {Schneider}, {Schreiber}, {Schwope}, {Sibley}, {Simmons},
  {Snedden}, {Allyn Smith}, {Stark}, {Stauffer}, {Steinmetz}, {Stoughton},
  {SubbaRao}, {Szalay}, {Szkody}, {Thakar}, {Sivarani}, {Tucker}, {Uomoto},
  {Vanden Berk}, {Vidrih}, {Wadadekar}, {Watters}, {Wilhelm}, {Wyse}, {Yarger},
  \& {Zucker}}]{2009Yanny}
{Yanny}, B., {Rockosi}, C., {Newberg}, H.~J., {et~al.} 2009, \aj, 137, 4377,
  \dodoi{10.1088/0004-6256/137/5/4377}

\bibitem[{{Yu} {et~al.}(2016){Yu}, {Diaferio}, {Agulli}, {Aguerri}, \&
  {Tozzi}}]{2016Yu}
{Yu}, H., {Diaferio}, A., {Agulli}, I., {Aguerri}, J.~A.~L., \& {Tozzi}, P.
  2016, \apj, 831, 156, \dodoi{10.3847/0004-637X/831/2/156}

\bibitem[{{Yu} {et~al.}(2015){Yu}, {Serra}, {Diaferio}, \& {Baldi}}]{Yu2015}
{Yu}, H., {Serra}, A.~L., {Diaferio}, A., \& {Baldi}, M. 2015, \apj, 810, 37,
  \dodoi{10.1088/0004-637X/810/1/37}

\bibitem[{{Zhao} \& {He}(1990)}]{1990Zhao}
{Zhao}, J.~L., \& {He}, Y.~P. 1990, \aap, 237, 54

\bibitem[{{Zhong} {et~al.}(2019){Zhong}, {Chen}, {Kouwenhoven}, {Li}, {Shao},
  \& {Hou}}]{2019Zhong}
{Zhong}, J., {Chen}, L., {Kouwenhoven}, M.~B.~N., {et~al.} 2019, \aap, 624,
  A34, \dodoi{10.1051/0004-6361/201834334}

\bibitem[{{Zinn} {et~al.}(2019){Zinn}, {Pinsonneault}, {Huber}, \&
  {Stello}}]{2019Zinn}
{Zinn}, J.~C., {Pinsonneault}, M.~H., {Huber}, D., \& {Stello}, D. 2019, \apj,
  878, 136, \dodoi{10.3847/1538-4357/ab1f66}

\end{thebibliography}

\end{CJK*}

\end{document}